\documentclass{iopjournal}
\usepackage[square,numbers]{natbib}
\usepackage{subcaption}
\usepackage{parskip}
\usepackage{amssymb}
\usepackage{bm}
\usepackage{amsmath}
\usepackage{float}
\newcommand{\tw}{t_{\text{w}}}
\newcommand{\chidyn}{\chi_{\text{st}}^{\text{dyn}}}
\begin{document}
%%%%%%%%%%%%%%%%%%%%%%%%%%%%%%%%%%%%%%%%%%%%%%%%%%%%%%%%%%%%%%%%%%%%%%%%%%%%%%%%
\articletype{Paper} %	 e.g. Paper, Letter, Topical Review...

\title{Unraveling anomalous relaxation effects in the thermodynamic limit}

\author{Emilio Pomares$^{1}$\orcid{0009-0007-5293-0268},
        V\'{\i}ctor Mart\'{\i}n-Mayor$^{1}$\orcid{0000-0002-3376-0327},
        Antonio Lasanta$^{2,3,4,*}$\orcid{0000-0003-4113-4358}
        and
        Gabriel \'Alvarez$^{1}$\orcid{0000-0003-2844-276X}}

\affil{$^1$Departamento de F\'{\i}sica Te\'orica, Facultad de Ciencias F\'{\i}sicas,
       Universidad Complutense de Madrid, 28040 Madrid, Spain}

\affil{$^2$Departamento de \'Algebra, Facultad de Educaci\'on, Econom\'{\i}a y Tecnolog\'{\i}a de Ceuta,
       Universidad de Granada, 51001 Ceuta, Spain}
       
\affil{$^3$Instituto Carlos I de F\'isica Te\'orica y Computacional, Universidad de Granada, E-18071 Granada, Spain}

\affil{$^4$Nanoparticles Trapping Laboratory, Universidad de Granada, Granada, Spain}

\affil{$^*$Author to whom any correspondence should be addressed.}

\email{alasanta@ugr.es}

%%%%%%%%%%%%%%%%%%%%%%%%%%%%%%%%%%%%%%%%%%%%%%%%%%%%%%%%%%%%%%%%%%%%%%%%%%%%%%%%
\begin{abstract}
We address two central open problems in the theory of anomalous Mpemba-like relaxations:
their extension beyond one spatial dimension and their consistent formulation in the thermodynamic limit.
Our framework is the antiferromagnetic Ising model on a square lattice under an externally applied magnetic field,
which enables us to work in the presence of a phase transition. The rich phase diagram contains
two control parameters: temperature and magnetic field. We demonstrate that the standard assumption of relaxation dominated
by a single leading exponential is inconsistent for intensive observables exhibiting standard fluctuations. Instead, as the system size increases,
a continuous spectrum of time scales emerges. Nevertheless, we make the ansatz that, in the vicinity of the phase transition,
the spectral projector onto the slowest time scales can be effectively characterized in terms of an equilibrium thermodynamic
quantity: the susceptibility associated with the order parameter of the metastable phase.
Combined with the richness of the phase diagram, this ansatz yields qualitative and semi-quantitative predictions
for optimal protocols leading to a variety of anomalous relaxation phenomena involving simultaneous variations of temperature and magnetic field.
These include direct and inverse Mpemba effects, cooling-heating asymmetries, and faster heating induced by precooling.
Careful Monte Carlo simulations validate our theoretical predictions. Furthermore, minimal post-optimization suffices
to convert our analytically guided protocols into fully optimal ones that display anomalous relaxations in their most pronounced form.
\end{abstract}
%%%%%%%%%%%%%%%%%%%%%%%%%%%%%%%%%%%%%%%%%%%%%%%%%%%%%%%%%%%%%%%%%%%%%%%%%%%%%%%%
\section{Introduction}
%%%%%%%%%%%%%%%%%%%%%%%%%%%%%%%%%%%%%%%%%%%%%%%%%%%%%%%%%%%%%%%%%%%%%%%%%%%%%%%%
Engineering and predicting relaxation processes far from equilibrium remains one of the central open challenges in statistical physics.
Despite major advances throughout the twentieth century, most theoretical tools---beyond specific cases such as dilute molecular gases---are limited to linear
response theory and fluctuation theorems~\cite{Nyquist1928,Zwanzig1965,kubo1966,onsager1931,onsager19312,marconi2008,evans20081,evans20082}.
These frameworks describe small deviations from equilibrium or provide constraints on nonequilibrium fluctuations, but they do not offer a general strategy
to manage transient regimes or optimize relaxation times between prescribed states in freely evolving systems. Recent progress in understanding
anomalous relaxation phenomena suggests that such optimization may nevertheless be possible~\cite{TB26,GL24}.

A paradigmatic example is the Mpemba effect~\cite{mpemba1969,jeng2006,TB26,LR17,lasanta2017,BaityJesi2019,carollo2021,kumar2020}.
Consider two identical systems differing only in their initial temperatures, both suddenly placed in contact with a colder thermal bath.
The Mpemba effect occurs when the initially hotter system cools faster than the initially colder one---even though the latter starts closer to equilibrium.
This counterintuitive behavior challenges the naive expectation that proximity to equilibrium guarantees faster relaxation.

Historically, the phenomenon was popularized by Mpemba and Osborne in 1969, although references to similar observations date back to Aristotle and early modern
thinkers~\cite{webster1923aristotle,bacon1900opus,bacon1878novum,descartes1637discours,descartes1638mersenne,barker1775the,black1775the}.
For decades, the effect was studied almost exclusively in the context of water freezing. Numerous mechanisms were proposed to explain the observations,
including evaporation, frost-induced thermal contact differences, solute concentration changes, dissolved gases, supercooling and nucleation variability,
convection, and hydrogen-bond rearrangements. Some of these mechanisms---such as evaporation and frost---provide clear and reproducible routes
to anomalous freezing times. Others, particularly those involving nucleation and supercooling, highlight the extreme sensitivity of freezing processes
to stochastic fluctuations and microscopic details.

From these historical studies, several lessons emerge. First, no single mechanism universally explains the Mpemba effect in water.
Second, experimental results are highly sensitive to container geometry, surface roughness, dissolved impurities, and sample size.
Third, freezing experiments are intrinsically difficult due to the interplay between convection, diffusion, and stochastic nucleation processes.
These complexities motivated a shift away from water-specific explanations towards a broader theoretical understanding of anomalous relaxation.

Beginning around 2017, independent theoretical works reframed the Mpemba effect within nonequilibrium statistical mechanics and introduced
the notion of the inverse Mpemba effect~\cite{lasanta2017,LR17,kumar2022inverse}. In Markovian systems~\cite{van1992stochastic},
the phenomenon can be understood through spectral decomposition of the generator of the dynamics. Relaxation toward equilibrium
can be expressed as a superposition of decaying modes. If the initial condition of the hotter system suppresses or cancels contributions
from the slowest-decaying modes, the system effectively evolves along faster modes, thereby reaching equilibrium sooner. In this sense,
the Mpemba effect is not paradoxical but reflects the geometric structure of relaxation in state space.

This spectral mechanism has been demonstrated theoretically and experimentally in classical Markovian systems as well as in open quantum systems.
In both contexts, anomalous relaxation arises from the interplay between eigenvalues and eigenvectors of the dynamical
generator~\cite{teza2023eigenvalue,teza2023relaxation,teza2022far,carollo2021,zhang2025observation,moroder2024thermodynamics,aharony2024inverse,ares2025quantum,beato2026relaxation}. In particular, strong Mpemba effects occur when the dominant slow timescale itself changes discontinuously
for special initial conditions, leading to exponential speedups relative to generic relaxation~\cite{klich2019}. In systems where spectral methods
are not directly applicable, alternative strategies for controlling relaxation have been identified using macroscopic observables.
Examples include energy nonequipartition in water~\cite{torrente2019,gijon2019}, specific kurtosis conditions in granular gases~\cite{lasanta2017},
and correlation length effects in spin glasses~\cite{BaityJesi2019}. These approaches exploit structural properties of the system's macroscopic state
to selectively enhance fast relaxation channels or suppress slow ones. Beyond single quenches, multi-quench protocols have also been proposed to accelerate 
relaxation~\cite{GR20,pemartin2021,GL24,ibanez2026cost,guery2022}. Preheating protocols can reduce the weight of slow modes prior to the final cooling
step~\cite{GR20,GL24,guery2022}. Near phase transitions, especially in systems with many degrees of freedom, transient heating or cooling
can leverage domain growth dynamics and timescale separation to achieve faster convergence~\cite{pemartin2021}.
Other control techniques have similarly demonstrated relaxation speedup in both theoretical and experimental settings\cite{chittari2023,ibanez2026cost}.

However, achieving speedup is not trivial. The Kovacs effect illustrates that naive two-step quenches can induce memory effects that hinder relaxation
rather than accelerate it. Systems can display nonmonotonic temporal responses in which macroscopic observables overshoot or undershoot their asymptotic
values~\cite{Kovacs1964,Kovacs1979,militaru2021,prados2014kovacs,Prados2014,mompo2021}. Thus, designing effective control protocols requires careful
understanding of mode structure and system geometry. Even more strikingly, recent studies have revealed a fundamental asymmetry between heating and cooling 
processes far from equilibrium~\cite{ibanez2023,tejero2025asymmetries}. Symmetric temperature changes do not necessarily produce symmetric relaxation dynamics. 
This asymmetry persists even for reciprocal relaxation processes between two fixed temperatures. The phenomenon has been successfully interpreted using thermal 
kinematics, a framework grounded in information geometry~\cite{Ito2020Stochastic,crooks2007,bravetti2025asymmetric}. Within this geometric picture,
relaxation trajectories in probability space possess directional structure that breaks naive symmetry between heating and cooling.

These developments motivate a unifying classification of anomalous relaxation phenomena. The term now encompasses a wide variety of nonmonotonic relaxation
behaviors across classical, granular, glassy, and quantum systems. Effects may be direct (cooling) or inverse (heating), equilibrium or steady-state,
thermal or non-thermal, weak or strong, classical or quantum. Relaxation time itself may be defined through phase-transition crossing,
final approach to equilibrium, or crossing of specific observables such as energy or effective temperature. This taxonomy clarifies that the anomalous
relaxation and Mpemba effect are not a single mechanism but a structural property of nonequilibrium dynamics~\cite{TB26}.

Therefore, the modern perspective on the Mpemba effect has renovated a historical puzzle about freezing water into a general principle
of nonequilibrium relaxation. Proximity to equilibrium does not uniquely determine relaxation speed; rather, the decomposition of initial
conditions into dynamical modes governs transient evolution. Through spectral analysis, macroscopic observables, and carefully
designed quench protocols, it becomes possible not only to explain anomalous relaxation but to harness it for controlling speedup. 

In conclusion, a common feature underlying anomalous relaxation phenomena is the nonmonotonic dependence of the relaxation time
on the initial probability distribution as the system evolves toward its final stationary state or undergoes a phase transition.
Although a clear theoretical picture has emerged for mean-field models, the behavior in 2D and 3D systems remains poorly understood.
The case of kinetic theory framework complements these approaches by enabling the analysis of anomalous relaxation and providing
a good quantitative description in the thermodynamic limit. Nevertheless, several fundamental questions remain unresolved.
In particular, the manifestation of the Mpemba effect in the thermodynamic limit, where vanishing spectral gaps and continuous spectra may arise,
remains largely unexplored. A further open issue concerns the identification of the physical observable that governs the anomalous effect
and whether this quantity is universal or system-dependent. In this manuscript, we focus on addressing some of these open questions.

Within this broad context, we present a theoretical approach focuses on controlling out-of-equilibrium evolution using only equilibrium physical observables
and the spectral decomposition of the dynamical generator.  The central physical idea is straightforward: by identifying the slowest-decaying physical observables,
one can prepare initial conditions that project the system onto faster-decaying modes, thereby accelerating total relaxation.
Near first-order phase transitions, where metastability and timescale separation are prominent, fast relaxation can be achieved
either by selecting suitable initial conditions or by briefly heating or cooling the system prior to final relaxation. Such strategies exploit the underlying
mode structure without requiring external forcing beyond simple quenches. To illustrate these ideas concretely, the framework were implemented
in the one-dimensional antiferromagnetic Ising model in a magnetic field. In such systems, spectral properties and observable projections
can be explicitly computed, providing a transparent demonstration of how anomalous relaxation and controlled speedup emerge~\cite{GL24}. 
In this paper, we extend previous results to a 2D system, namely the antiferromagnetic Ising model, in the thermodynamic limit.

The content of the remaining part of this work is organized as follows. Section~\ref{sect:model-and-dynamics} describes our framework.
This includes a description of our model and main observables (section~\ref{subsect:model-observables}) and of the phase diagram
(section~\ref{subsec:phase_diagram}). Our model dynamics is discussed in section~\ref{subsect:dynamics}. We also address  the approach
to the thermodynamic limit from both the viewpoint of the equilibrium mean values (section~\ref{subsect:TL-statics})
and of the  dynamics (section~\ref{subsect:TLdynamics}). In light of our general discussion about the thermodynamic limit,
we put forth in section~\ref{sec:effective_description} an effective description of the dynamics. These considerations crystallize in our main conjecture,
see equation~\eqref{eq:our-main-hyp} in section~\ref{sec:anomalous_relaxation}. Also in section~\ref{sec:anomalous_relaxation}
we show how the conjecture allows us to successfully predict a large variety of Mpemba-like anomalous relaxations.
Finally, the paper ends in section~\ref{sec:conclusions} with some  brief comments on possible extensions and limitations of our approach.
%%%%%%%%%%%%%%%%%%%%%%%%%%%%%%%%%%%%%%%%%%%%%%%%%%%%%%%%%%%%%%%%%%%%%%%%%%%%%%%%
\section{The 2D antiferromagnetic Ising model on a square lattice}\label{sect:model-and-dynamics}
%%%%%%%%%%%%%%%%%%%%%%%%%%%%%%%%%%%%%%%%%%%%%%%%%%%%%%%%%%%%%%%%%%%%%%%%%%%%%%%%
\subsection{Model definition and main observables}\label{subsect:model-observables}
%%%%%%%%%%%%%%%%%%%%%%%%%%%%%%%%%%%%%%%%%%%%%%%%%%%%%%%%%%%%%%%%%%%%%%%%%%%%%%%%
We consider an antiferromagnetic Ising model on a $2D$ square lattice with
$N=L\times L$ classical spins $\sigma_i=\pm 1$, and denote by $\Omega=\{1,-1\}^N$
the set of all possible states $\mathbf{x}=\{\sigma_1,\ldots,\sigma_N\}$ of the system.

The energy $\mathcal{E}$ of a given state $\mathbf{x}$ has two contributions, the exchange energy $\mathcal{E}_J$
and the Zeeman energy $\mathcal{E}_h$,
\begin{equation}\label{eq:total_energy}
        \mathcal{E}(\mathbf{x})= \mathcal{E}_J(\mathbf{x})\ +\ \mathcal{E}_h(\mathbf{x})\,.
\end{equation}
The exchange energy is
\begin{equation}
    \mathcal{E}_J(\mathbf{x}) = -J\sum_{\langle i,j \rangle} \sigma_i \sigma_j,
\end{equation}
where $\langle i, j \rangle$ denotes summation over nearest neighbors in the square
lattice (with periodic boundary conditions), and the negative coupling constant $J<0$
qualifies the interaction as antiferromagnetic. The Zeeman energy is defined in terms
of the total (uniform) magnetization
\begin{equation}\label{eq:magnetization}
    \mathcal{M_{\mathrm{u}}}(\mathbf{x}) = \sum_{i=0}^{N-1} \sigma_i,
\end{equation}
as
\begin{equation}
    \mathcal{E}_h(\mathbf{x}) = - h\, \mathcal{M_{\mathrm{u}}}(\mathbf{x}), 
\end{equation}
where the external magnetic field is $h>0$. 

The partition function at the thermal bath temperature $T_b$ is given by
\begin{equation}
    Z_N(T_b) = \sum_{\mathbf{x}\in\Omega} \exp\left(-\mathcal{E}(\mathbf{x})/k_B T_b\right),
\end{equation}
and the corresponding Boltzmann weight for the state $\mathbf{x}$ by
\begin{equation}
    \pi_{\mathbf{x}}^{T_b} =\frac{\exp\left(-\mathcal{E}(\mathbf{x})/k_B T_b\right)}{ Z_N(T_b)}.
\end{equation}
Therefore, the equilibrium expected value of an observable $\mathcal{A}$ at
temperature $T$ is
\begin{equation}
    \mathbb{E}^T[\mathcal{A}]
    =
    \sum_{\mathbf{x}\in\Omega} \pi_{\mathbf{x}}^T \mathcal{A}(\mathbf{x})\,.
\end{equation}
We shall be working in units such that $k_B=-J=1$.
We remark that in reference~\cite{GR20} $J$ is replaced by $J/4$, and ultimately set to $1$.
Therefore, when comparing with reference~\cite{GR20}, our $h$ will have to be scaled by $4$.

It will be also important to quantify the length scale that characterize the correlations in the antiferromagnetic fluctuations.
To do so, we introduce the wave vector dependent staggered magnetization $\widetilde{\mathcal{M}}_\mathrm{st}^{(k_x,k_y)}$
and its associated susceptibility $F_{\text{st}}(k_x,k_y)$,
\begin{equation}
    \widetilde{\mathcal{M}}_\mathrm{st}^{(k_x,k_y)}(\mathbf{x}) = \sum_{i=0}^{N-1} (-1)^{x_i+y_i} \,\text{e}^{\mathrm{i}(k_x x_i+k_y y_i)}\,\sigma_i,\quad F_{\text{st}}(k_x,k_y)=\frac{1}{N}\mathbb{E}^{T_b}\Big[\big(\widetilde{\mathcal{M}}_\mathrm{st}^{(k_x,k_y)}\big)^2\Big]\,,
\end{equation}
where $x_i$ and $y_i$ denote the Cartesian coordinates in the square lattice of the
spin $i$ (we are using the lexicographic---or row-minor---ordering $i=x_i+L y_i$, with $x_i,y_i=0,1,\ldots ,L-1$). 
Two particular cases are of importance: the uniform magnetization $\mathcal{M}_{\text{u}}=\widetilde{\mathcal{M}}_\mathrm{st}^{(\pi,\pi)}$
defined in equation~(\ref{eq:magnetization}), and the staggered magnetization
\begin{equation}
    \mathcal{M}_\mathrm{st}(\mathbf{x})
    =
    \widetilde{\mathcal{M}}_\mathrm{st}^{(0,0)}(\mathbf{x})
    =
    \sum_{i=0}^{N-1} (-1)^{x_i+y_i} \sigma_i\,.
\end{equation}
Notice that if the spin configuration is shifted by one lattice unit in either the $x$- or $y$-direction,
the staggered magnetization changes sign, $\mathcal{M}_\mathrm{st}\longrightarrow -\mathcal{M}_\mathrm{st}$,
while both $\mathcal{E}_J$ and $\mathcal{M}_{\text{u}}$ (and hence $\pi^{T_b}$) remain unchanged. It follows that 
\begin{equation}
    \mathbb{E}^{T_b}[\mathcal{M}_{\text{st}}]=0\,,
\end{equation}
(at least for any finite lattice size $L$). Yet, when a staggered interaction  $-h_{\text{st}} \mathcal{M}_\mathrm{st}$
is added to the energy, a non-vanishing exchange magnetization arises, which can be quantified by the so called staggered susceptibility
\begin{equation}\label{eq:stagger-suscept-def}
\chi_\mathrm{st}=\frac{T_b}{N} \frac{\partial\mathbb{E}^{T_b}[\mathcal{M}_{\text{st}}]}{\partial h_{\text{st}}}\,,\quad
    \chi_\mathrm{st}(T_b,h_{\text{st}}=0) = F_{\text{st}}(0,0)=\frac{1}{N} \mathbb{E}^{T_b}[\mathcal{M}_\mathrm{st}^2]\,.
\end{equation}
As the factor $1/N$ in the definition of the staggered susceptibility suggests, we shall use
the intensive (or per spin) version of the different observables ($\mathcal{M}_{\text{st}}/{N}$, for instance)
to compare results obtained for different lattice sizes. 

Finally, since we shall be working with $h_{\text{st}}=0$ (see, for instance, reference~\cite{AM:05}),  we will compute
the second-moment staggered correlation length $\xi_{\text{st}}$ from the staggered
susceptibilities at zero wave vector and at the minimal wave vector compatible with the periodic boundary
conditions $F_{\text{st}}(2\pi/L,0)=F_{\text{st}}(0,2\pi/L)$ as
\begin{equation}\label{eq:stagger-xi-def}
    \xi_{\text{st}}=\frac{1}{2\sin(\pi/L)}\sqrt{\frac{\chi_{\text{st}}}{F_{\text{st}}(2\pi/L,0)}\ -\ 1}\,.
\end{equation}
%%%%%%%%%%%%%%%%%%%%%%%%%%%%%%%%%%%%%%%%%%%%%%%%%%%%%%%%%%%%%%%%%%%%%%%%%%%%%%%%
\subsection{Phase diagram}\label{subsec:phase_diagram}
%%%%%%%%%%%%%%%%%%%%%%%%%%%%%%%%%%%%%%%%%%%%%%%%%%%%%%%%%%%%%%%%%%%%%%%%%%%%%%%%
%%%%%%%%%%%%%%%%%%%%%%%%%%%%%%%%%%%%%%%%%%%%%%%%%%%%%%%%%%%%%%%%%%%%%%%%%%%%%%%%
\begin{figure}[t]
    \centering
    \includegraphics[width=0.48\linewidth]{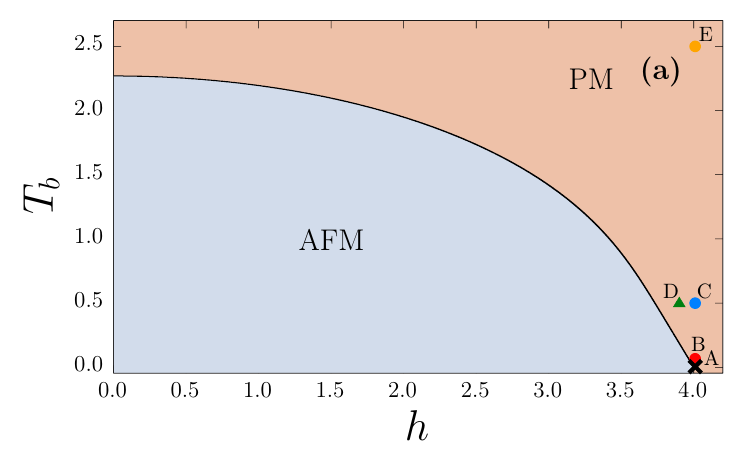}
    \includegraphics[width=0.48\linewidth]{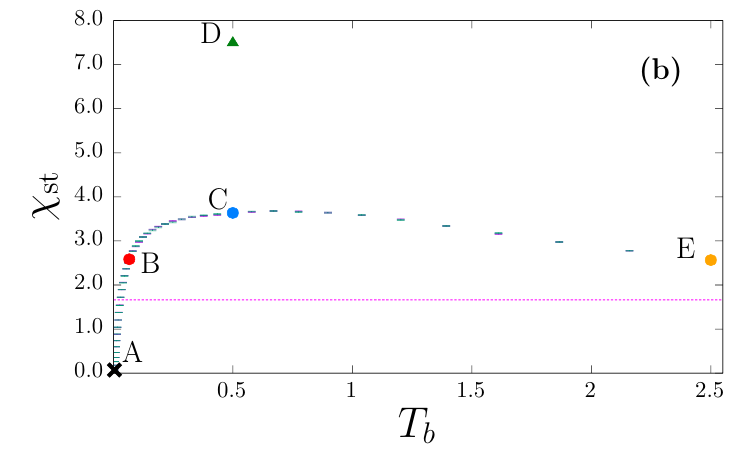}
    \caption{(a) Phase map of our 2D AFM Ising model showing the critical line from reference~\cite{WK97}.
                       Our working points A to E are summarized in table~\ref{tab:working_points}. (b) $\chi_{\mathrm{st}}$ as a function of the temperature
                       $T_b$ for lattices $N = 256\times256$ (purple)  and $N = 512\times512$ (green), which appear almost superimposed.
                       For points A, B, C and E, $h=4.01$ (point A has a cross marker to aid visibility). For point D (triangle marker), $h=3.9$.
                       The horizontal dashed line shows the upper bound on $\chi_{\mathrm{st}}$ found for the 1D model.}
    \label{fig:crit-suscept}
\end{figure}
%%%%%%%%%%%%%%%%%%%%%%%%%%%%%%%%%%%%%%%%%%%%%%%%%%%%%%%%%%%%%%%%%%%%%%%%%%%%%%%%

In figure~\ref{fig:crit-suscept}(a) we show the phase map for our 2D antiferromagnetic Ising model with a magnetic field on
the square lattice~\eqref{eq:total_energy}, including an analytic approximation to the critical line taken from reference~\cite{WK97}.
The critical line separates the paramagnetic phase, where $\mathbb{E}^{T_b}[\mathcal{M}_\text{u}]>0$ and $\mathbb{E}^{T_b}[\mathcal{M}_{\text{st}}]=0$,
from the antiferromagnetic phase, where an infinitesimal staggered field $h_\text{st}=0^+$ induces positive values for both magnetizations
$\mathbb{E}^{T_b}[\mathcal{M}_\text{u}]>0$ and $\mathbb{E}^{T_b}[\mathcal{M}_{\text{st}}]>0$. The critical line ends at the critical point at $h=0$
and $T=2/\log(1+\sqrt{2})$, where the correlation length $\xi_\text{st}$ diverges. This divergence implies scaling behaviors
for the relaxation time $\tau$ and the staggered susceptibility (see, for instance, reference~\cite{ZJ05})
\begin{equation}\label{eq:tau_xist_dyn_exp}
    \tau \sim (\xi_\mathrm{st})^{z}\,,\quad \chi_\text{st}\sim  (\xi_\mathrm{st})^{\gamma/\nu}\,,
\end{equation}
where the critical exponents for the 2D Ising Model universality class are  $z=2.1676(1)$~\cite{VA25} and $\gamma/\nu=7/4$.
This behavior is in sharp contrast with the 1D case, where $\chi_{\text{st}}$ is bounded above by 1.663~\cite{GL24}.
Hence, in the 2D model we have access to arbitrarily larger values of $\xi_{\text{st}}$ which entails arbitrarily long relaxation times.
This key difference leads us to anticipate more striking manifestations of anomalous relaxations for the 2D model.

We also mark in figure~\ref{fig:crit-suscept}(a)  points $\text{A}$ to $\text{E}$ that we have selected to exhibit these anomalous relaxations.
The corresponding numerical values  are given in table~\ref{tab:working_points}. Points $A$, $C$, and $E$ are taken, after suitable rescaling,
from reference~\cite{GR20}. Taking advantage of the scaling relation for $\tau$ in equation~\eqref{eq:tau_xist_dyn_exp},
we select two additional points ($B$ and $E$) with relaxation times larger than those achievable in the $1D$ model. 

%%%%%%%%%%%%%%%%%%%%%%%%%%%%%%%%%%%%%%%%%%%%%%%%%%%%%%%%%%%%%%%%%%%%%%%%%%%%%%%%
\begin{table}
\caption{Phase map coordinates of points $\text{A}$, $\text{B}$, $\text{C}$, $\text{D}$,~and $\text{E}$
         in figure~\ref{fig:crit-suscept}.}
\centering
\begin{tabular}{ccc}
\hline
Label & $T_b$ & $h$\\
\hline
$\text{A}$ & 0.005 & 4.01 \\
$\text{B}$ & 0.067 & 4.01 \\
$\text{C}$ & 0.5 & 4.01 \\
$\text{D}$ & 0.5 & 3.9 \\
$\text{E}$ & 2.5 & 4.01 \\
\hline
\end{tabular}
\label{tab:working_points}
\end{table}
%%%%%%%%%%%%%%%%%%%%%%%%%%%%%%%%%%%%%%%%%%%%%%%%%%%%%%%%%%%%%%%%%%%%%%%%%%%%%%%%

In figure~\ref{fig:crit-suscept}(b)  we show the staggered susceptibility $\chi_\mathrm{st}$ as a function of $T_b$,  for $h=4.01$
and lattices of sizes $256\times 256$ (purple) and $512\times 512$ (green). The two values appear almost superimposed
and the error bars are barely noticeable (further evidence that we have effectively reached the thermodynamic limit is provided
in section~\ref{subsect:TL-statics}). The key feature is the non monotonic behavior of $\chi_\text{st}$ as a function of $T_b$,
a feature that survives the thermodynamic limit. In $1D$, the  observable displaying the slowest relaxation
to equilibrium---see section~\ref{subsect:dynamics}---was shown to have a large projection onto $\chi_\text{st}$.
Although the thermodynamic limit complicates the analysis of the relaxation---see section~\ref{subsect:TLdynamics}---one may hope
that the group of most relevant excitations still have large projections onto  $\chi_\text{st}$.
We have verified this expectation by designing from figure~\ref{fig:crit-suscept}(b) thermal protocols that exhibit anomalous relaxations,
e.g., heating-cooling asymmetries, accelerated heatings, and both direct and inverse Mpemba effects.

As we mentioned earlier, point $D$ has been chosen as a more refined representative of effects that are not achievable in 1D.
In fact, the corresponding value of $\chi_{\text{st}}$ (and therefore of $\xi_{\text{st}}$), is significantly higher than the maximum
for the curve~\ref{fig:crit-suscept}(b), where the magnetic field is kept fixed and only the temperature is varied.
While by approaching the critical point at $h=0$ we could have selected points in the phase diagram with arbitrarily large values of
$\chi_{\text{st}}$ (and, hence, of $\xi_{\text{st}}$), we have compromised to avoid prohibitive computational times.
Indeed, the simulation time should be proportional to the relaxation time, which is approximately 2.5 times longer for point D
than for point C: $\tau(\text{D})/\tau(\text{C})\approx \big(\xi_{\text{st}}(\text{D})/\xi_{\text{st}}(\text{C})\big)^{z} \approx 2.5$.
%%%%%%%%%%%%%%%%%%%%%%%%%%%%%%%%%%%%%%%%%%%%%%%%%%%%%%%%%%%%%%%%%%%%%%%%%%%%%%%%
\subsection{The dynamical algorithm}\label{subsect:dynamics}
%%%%%%%%%%%%%%%%%%%%%%%%%%%%%%%%%%%%%%%%%%%%%%%%%%%%%%%%%%%%%%%%%%%%%%%%%%%%%%%%
In condensed matter experiments time is a continuous variable. Hence, we have preferred to model the dynamics as a continuous-time Markov chain,
which is obtained by taking the continuum limit of a discrete-time dynamics~\cite{LP17}. We have used the standard heat-bath algorithm with random access
to the lattice~\cite{SO97} starting from the random-access version of the discrete algorithm to preserve its detailed-balance property
when taking the continuous time limit. We summarize here only the basic steps and refer for details to reference~\cite{LP17}.
%%%%%%%%%%%%%%%%%%%%%%%%%%%%%%%%%%%%%%%%%%%%%%%%%%%%%%%%%%%%%%%%%%%%%%%%%%%%%%%%
\subsubsection{The initial discrete-time dynamics}
%%%%%%%%%%%%%%%%%%%%%%%%%%%%%%%%%%%%%%%%%%%%%%%%%%%%%%%%%%%%%%%%%%%%%%%%%%%%%%%%
The discrete time dynamics is described by its generating matrix $\textbf{T}$, whose matrix element $ \textbf{T}_{\mathbf{x},\mathbf{y}}$ represents
the probability of reaching state $\mathbf{y}$ in a single time step, conditional to having $\mathbf{x}$ as initial configuration.
This matrix element is sometimes more pictorially written as $\textbf{T}\mathbf{x}\to\mathbf{y}$.
As every probability, $\textbf{T}$ is non-negative and well normalized
\begin{equation}\label{eq:completitud}
    \textbf{T}_{\mathbf{x},\mathbf{y}} \geq 0\,,
    \quad
    \sum_{\mathbf{y}\in\Omega} \,\textbf{T}_{\mathbf{x},\mathbf{y}}=1\,.
\end{equation}
More specifically, $\textbf{T}$ describes a single spin-flip dynamics, wherein the only non-vanishing matrix elements are those
in which the initial and final configurations differ at most in the value of one spin.
In the case where $\mathbf{x}$ and $\mathbf{y}$ are related by exactly one spin flip we write,
\begin{eqnarray}\label{eq:heat-bath}
    \textbf{T}_{\mathbf{x},\mathbf{y}} & = & \frac{1}{N} \, R^\text{HB}_{\mathbf{x},\mathbf{y}}\,,
    \\
    R^\text{HB}_{\mathbf{x},\mathbf{y}} & = &
    \frac{ \exp \left( -\left(\mathcal{E}(\boldsymbol{y})-\mathcal{E}(\boldsymbol{x}) \right) /T_{b} \right) }{
    1+\exp \left( -\left(\mathcal{E}(\boldsymbol{y})-\mathcal{E}(\boldsymbol{x}) \right) /T_{b} \right)},\label{eq:heat-bath-accept}
\end{eqnarray}
(recall that we work in $k_{\text{B}}=1$ units).

The matrix $\textbf{T}$ acts on two different vector spaces, both of them isomorphic to the space $\mathcal{F}(\Omega)$
of real-valued functions on $\Omega$. The probability functions for the spin configurations belong to the first of these vector spaces,
while the second space is the space of observables. We shall denote by $\pi^{T_b}$, $\mathcal{M}_\mathrm{u}$, etc., elements
of $\mathcal{F}(\Omega)$, whereas $\pi^{T_b}_\mathbf{x}$, $\mathcal{M}_\mathrm{u}(\mathbf{x})$, etc., will denote their
respective values for the spin configuration $\mathbf{x}$.

The strong form of the master equation gives the probability over $\Omega$ of finding the system in state $\mathbf{x}$
at the discrete time $s+n$, $P_{s+n}(\mathbf{x})$:
\begin{equation}\label{eq:Master}
P_{s+n}(\mathbf{x})=\sum_{\mathbf{y}\in\Omega}\, P_s(\mathbf{y}) [\textbf{T}^n]_{\mathbf{y},\mathbf{x}}\,,
\end{equation}
where $\textbf{T}^n$ is the $n$-th power of matrix $\textbf{T}$ and $P_s$ is the probability at the discrete time $s$.
Note that $P$ is a row-vector and the matrix $\textbf{T}$ acts by left-multiplication. Conversely, the weak form of the Master equation
gives the time evolution of the observables through right-multiplication,
\begin{equation}\label{eq:Master-debil}
\mathcal{T}^n[\mathcal{A}](\mathbf{x})=\sum_{\mathbf{y}\in\Omega}\,  [\textbf{T}^n]_{\mathbf{x},\mathbf{y}} \mathcal{A}(\mathbf{y})\,.
\end{equation}
Note that in effect $\mathcal{T}^n[\mathcal{A}]$ is a new observable whose value in state $\mathbf{x}$ is the expectation
value for the observable $\mathcal{A}$  at the discrete time $s+n$ of a system that was in state $\mathbf{x}$ at the discrete time $s$. 

The detailed balance condition is equivalent to the self-adjointness of the operator $\cal{T}$ for the scalar product of
two operators~\cite{SO97} 
\begin{equation}
    \label{eq:producto_escalar}
    \langle\mathcal{A}\,|\,\mathcal{B}\rangle
    =
    \mathbb{E}^{T_b}[\mathcal{A} \, \mathcal{B}]
    =
    \sum_{\mathbf{x} \in \Omega}
        \pi_{\mathbf{x}}^{T_b} \mathcal{A}(\mathbf{x})\mathcal{B}(\mathbf{x}).
\end{equation}
Since $\Omega$ is finite, the vector space $\mathcal{F}(\Omega)$ is finite-dimensional (in fact, of dimension $2^N$)
and the spectral theorem of linear algebra implies that $\mathcal{F}(\Omega)$ has an orthogonal basis of right-eigenvectors
of the selfadjoint operator $\mathcal{T}$. The eigenvalues $\Lambda_i, i=1,2,\ldots, 2^N$ are real,
and due to equation~\eqref{eq:completitud}, $|\Lambda_i|\leq 1$. It follows that matrix $\textbf{T}$ has an orthogonal basis of left eigenvectors
(the spectrum of $\textbf{T}$ and $\mathcal{T}$ are identical because the two matrices are transposed of one another).

The strong and weak form of the master equation are formally solved in terms of their respective eigenvectors.
We arrange the eigenvalues in decreasing order,
\begin{equation}
\Lambda_1=1 > \Lambda_2\geq \Lambda_2\geq\ldots\,,
\end{equation}
and denote the respective eigenvectors by
\begin{eqnarray}\label{eq:eigenvectors_left_def}
v_k \textbf{T}&=& \Lambda_k v_k\,,\quad v_1=\pi^{T_b}\,,\\
\mathcal{T}[O_k]&=&\Lambda_k O_k\,,\quad O_1=\mathbf{1}\,,\label{eq:eigenvectors_right_def}
\end{eqnarray}
where $\mathbf{1}$ is the constant observable that equals 1 for all configurations ($\mathbf{1}(X)=1\,\forall \mathbf{x}$),
and in particular $\mathbb{E}^{T_b}[\mathcal{A}]=\langle\mathbf{1}|\mathcal{A}\rangle$.

Hence, the formal solution of the Master equation is
\begin{equation}\label{eq:strong-sol-discrete}
P_{n=0}=\pi^{T_b}\ +\ \sum_{k>1}\,\gamma_k v_k\,,\quad 
P_{n}=\pi^{T_b}\ +\ \sum_{k>1}\,  \Lambda_k^n\, \gamma_k v_k\,,
\end{equation}
while the dynamic evolution of an observable $\mathcal{A}$ is
\begin{eqnarray}\label{eq:weak-sol-discrete}
\mathcal{T}^n[A]&=& \mathbb{E}^{T_b}[\mathcal{A}]\mathbf{1}\ +\ \sum_{k>1}\, \Lambda_k^n\, \alpha_k O_k\,,\\
\alpha_k&=& \langle O_k|\mathcal{A}\rangle.\label{eq:alpha_coef_def}
\end{eqnarray}
Hence, because $|\Lambda_k|<1$ if $k>1$, the expectation value for $\mathcal{A}$ approaches $\mathbb{E}^{T_b}[\mathcal{A}]$
in the large $n$ limit, independently of the initial configuration. Hereafter we will denote with a superscript ${}^\perp$
the fluctuating part of an observable, e.g.,
\begin{equation}
 \mathcal{A}^\perp=\mathcal{A}-\mathbb{E}^{T_b}[\mathcal{A}]\mathbf{1}.
\end{equation}
%%%%%%%%%%%%%%%%%%%%%%%%%%%%%%%%%%%%%%%%%%%%%%%%%%%%%%%%%%%%%%%%%%%%%%%%%%%%%%%%
\subsubsection{The continuous time limit}
%%%%%%%%%%%%%%%%%%%%%%%%%%%%%%%%%%%%%%%%%%%%%%%%%%%%%%%%%%%%%%%%%%%%%%%%%%%%%%%%
The key idea  of the limiting process~\cite{LP17} will be to preserve the form of equations~(\ref{eq:completitud}), 
( \ref{eq:strong-sol-discrete}), (\ref{eq:weak-sol-discrete}), (\ref{eq:alpha_coef_def}) and the left- and right-eigenvectors~(\ref{eq:eigenvectors_left_def}),
and (\ref{eq:eigenvectors_right_def}). Only the eigenvalues will differ from the discrete case.

The basic quantity is the matrix $R^{\text{HB}}$, whose non-diagonal matrix elements are in equation~\eqref{eq:heat-bath-accept}.
Note that $\textbf{T}$ and $R^{\text{HB}}$ are related as $R^{\text{HB}}= N (\textbf{T}-\mathbf{I})$, where $\mathbf{I}$ is the identity matrix.
Hence the diagonal elements of $R^{\text{HB}}$ are negative and given by
\begin{equation}
    \label{eq:completitud_2}
    R^{\text{HB}}_{\mathbf{x},\mathbf{x}}
    =
    - \sum_{\mathbf{y}\in\Omega\setminus\{\mathbf{x}\}}\, R^{\text{HB}}_{\mathbf{x},\mathbf{y}}\,.
\end{equation}
The eigenvalues for $R^{\text{HB}}$ are
\begin{equation}\label{eq:eigenvalue-ordered}
    \lambda_1=0>\lambda_1\geq\lambda_2\ldots\,,\quad \lambda_{2^N}\geq -2N\,.
\end{equation}

Then, a simple argument~\cite{LP17,GL24} allows us to go from the discrete time $n$ to a continuous time $t/t^{*}$, where $t^{*}$ is our time unit.
The formal solution of the continuous time evolution for the probability is given by
\begin{equation}\label{eq:formal-strong-solution}
P_{t=0}=\pi^{T_b}\ +\ \sum_{k>1}\,\gamma_k v_k\,,\quad 
P_{t/t^*}=\pi^{T_b}\ +\ \sum_{k>1}\, \text{e}^{-|\lambda_k|t/t^{*}}\, \gamma_k v_k\,,
\end{equation}
whereas for the observables we have
\begin{equation}
\mathcal{T}_{t/t^{*}}[A] = \mathbb{E}^{T_b}[\mathcal{A}]\mathbf{1}\ +\ \sum_{k>1}\, \text{e}^{-|\lambda_k|t/t^{*}}\, \alpha_k O_k\,.
\end{equation}

Although in the sequel more sophisticated preparation protocols will be used, a particular case of the previous equation
will be used so frequently that deserves explicit mention. If a system is brought to thermal equilibrium at a temperature $T^*$
with an expectation value for the observable $\mathcal{A}$ given by
\begin{equation}
    \label{eq:prep}
    E_{t=0}[\mathcal{A}] = \sum_\mathbf{y} \pi^{T^*}_\mathbf{y} \mathcal{A}(\mathbf{y})\,,
\end{equation}
and then suddenly placed at time $t=0$ in a thermal bath at temperature $T_b$, then the expectation value
for $\mathcal{A}$ at a later time $t>0$ is
\begin{equation}\label{eq:A-continuos-time}
E_{t}[\mathcal{A}]=\mathbb{E}^{T_b}[\mathcal{A}]\ +\ \sum_{k>1}\, \text{e}^{-|\lambda_k|t/t^{*}}\, \alpha_k \beta_{k}^{(O)}\,
\end{equation}
where
\begin{equation}
    \beta_{k}^{(O)}=E_{t=0}[O_k]\,.
\end{equation}

Although obtaining the spectral decomposition for the matrix $R^{\text{HB}}$ in our 2D case is unfeasible
(recall that the dimension of the configuration space $\Omega$ is $2^{L\times L}$), the previous discussion
leads to a simple algorithm for computer simulation, because the probability for the system to remain
in its current state $\mathbf{x}$ in the time interval from $t_1$ to $t$ is
\begin{equation}
P_\text{No change}(t,t_1)=\text{e}^{R^{\text{HB}}_{\mathbf{x},\mathbf{x}}(t-t_1)}\,,
\end{equation}
and this is a Poisson process ideally suited for an $n$-fold way simulation~\cite{BK75,GI77}.
We refer to reference~\cite{GL24} for implementation details. (The $n$ in $n$-fold should
not to be confused with the number of discrete time steps $n$.)
%%%%%%%%%%%%%%%%%%%%%%%%%%%%%%%%%%%%%%%%%%%%%%%%%%%%%%%%%%%%%%%%%%%%%%%%%%%%%%%%
\subsection{Thermodynamic limit, static}\label{subsect:TL-statics}
%%%%%%%%%%%%%%%%%%%%%%%%%%%%%%%%%%%%%%%%%%%%%%%%%%%%%%%%%%%%%%%%%%%%%%%%%%%%%%%%
We bring the system to thermal equilibrium using standard discrete-time heat bath dynamics~\cite{SO97} with the single-spin update rule
\begin{equation}
    p(\sigma_i \to -\sigma_i) = \frac{1}{1+e^{\Delta \mathcal{E}_i/T}},
\end{equation}
where $\Delta \mathcal{E}_i$ is the energy difference between the flipped and original configurations.

For performance reasons, we have used not random but sequential access to the lattice. Therefore, detailed balance is not satisfied
but the transition matrix does satisfy the weak global balance condition, which is sufficient to guarantee convergence
to the correct Gibbs distribution for self-averaging quantities such as the magnetization per spin.
Also for computational economy, we have used the same dynamics for the initial thermalization step
of some of the temperature-changing protocols, switching to our continuous-time algorithm afterwards.

\begin{figure}
    \begin{center}
        \includegraphics[width=8cm]{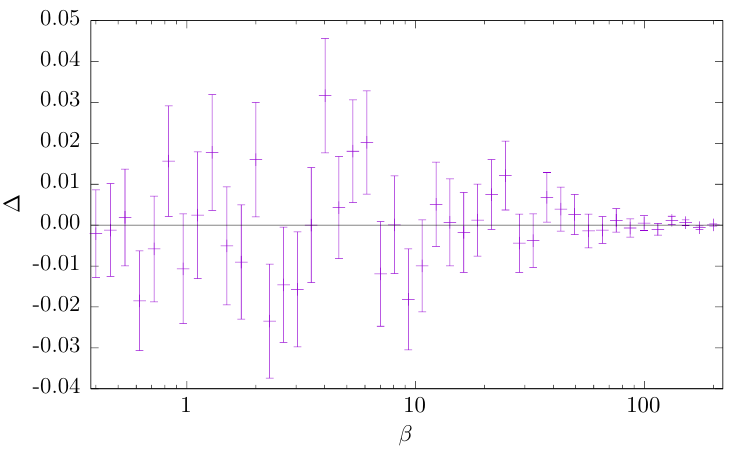}
    \end{center}
    \caption{Difference $\Delta=\chi_{\mathrm{st}}(L=256) - \chi_{\mathrm{st}}(L=512)$ as a function of the temperature $T$.\label{fig:diff_stg_suscept}}
\end{figure}

Figure~\ref{fig:diff_stg_suscept} shows that the difference in $\chi_{\mathrm{st}}$ as a function of
the temperature $T$ for two different lattice sizes ($N=256 \times 256$ and $N=512 \times 512$) is smaller than the simulation error.
Table~\ref{tab:xist} shows that the values of $\xi_{\mathrm{st}}$ already stabilize for a lattice of size $N = 32 \times 32$,
providing further evidence that we have effectively reached the thermodynamic limit in the static equilibrium regime.

%%%%%%%%%%%%%%%%%%%%%%%%%%%%%%%%%%%%%%%%%%%%%%%%%%%%%%%%%%%%%%%%%%%%%%%%%%%%%%%%
%%%%%%%%%%%%%%%%%%%%%%%%%%%%%%%%%%%%%%%%%%%%%%%%%%%%%%%%%%%%%%%%%%%%%%%%%%%%%%%%

%%%%%%%%%%%%%%%%%%%%%%%%%%%%%%%%%%%%%%%%%%%%%%%%%%%%%%%%%%%%%%%%%%%%%%%%%%%%%%%%
\begin{table}
\caption{Staggered correlation length $\xi_{\mathrm{st}}$ as a function of the lattice size $N=L\times L$ for the values of the
              temperature $T$ corresponding to points $\text{A}$, $\text{B}$, $\text{C}$, $\text{D}$~and $\text{E}$ in figure~\ref{fig:crit-suscept}.}
\centering
\begin{tabular}{cccccc}
\hline
Lattice size & $\xi_{\mathrm{st}}(\text{A})$
                   & $\xi_{\mathrm{st}}(\text{B})$
                   & $\xi_{\mathrm{st}}(\text{C})$
                   & $\xi_{\mathrm{st}}(\text{D})$
                   & $\xi_{\mathrm{st}}(\text{E})$ \\
\hline
$8\times 8$ & 0.131(4) & 1.041(1) & 1.286(1) & 1.952(2) & 0.901(1)\\
$16\times 16$ & 0.128(14) & 1.041(4) & 1.287(4) & 1.983(5) & 0.902(3)\\
$32\times 32$ & 0.114(59) & 1.051(12) & 1.289(13) & 1.994(15) & 0.909(11)\\
\hline
\end{tabular}
\label{tab:xist}
\end{table}
%%%%%%%%%%%%%%%%%%%%%%%%%%%%%%%%%%%%%%%%%%%%%%%%%%%%%%%%%%%%%%%%%%%%%%%%%%%%%%%%

%%%%%%%%%%%%%%%%%%%%%%%%%%%%%%%%%%%%%%%%%%%%%%%%%%%%%%%%%%%%%%%%%%%%%%%%%%%%%%%%
\subsection{Thermodynamic limit, dynamic}\label{subsect:TLdynamics}
%%%%%%%%%%%%%%%%%%%%%%%%%%%%%%%%%%%%%%%%%%%%%%%%%%%%%%%%%%%%%%%%%%%%%%%%%%%%%%%%
In this paragraph we give numerical evidence that we also reach the thermodynamic limit in the dynamic regime.
The theoretical framework will be addressed in section~\ref{sec:effective_description}.

\begin{figure}[ht]
    \centering
    \includegraphics[width=0.45\linewidth]{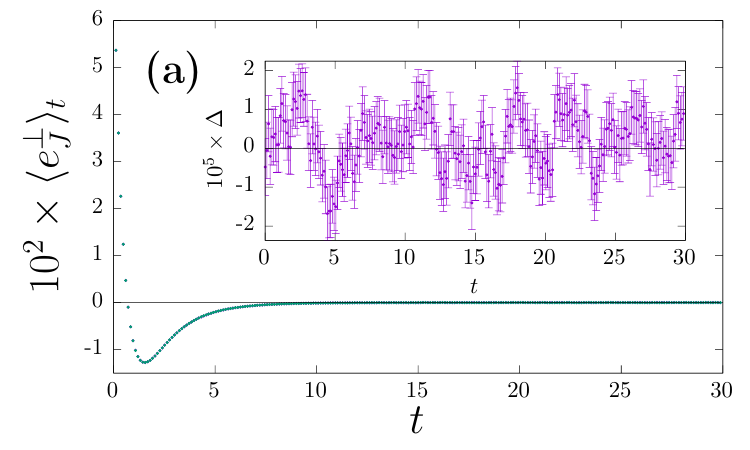}
    \includegraphics[width=0.45\linewidth]{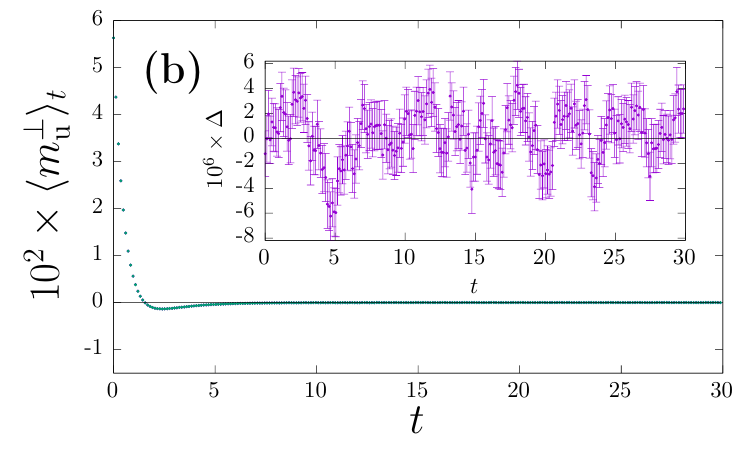}
    \includegraphics[width=0.45\linewidth]{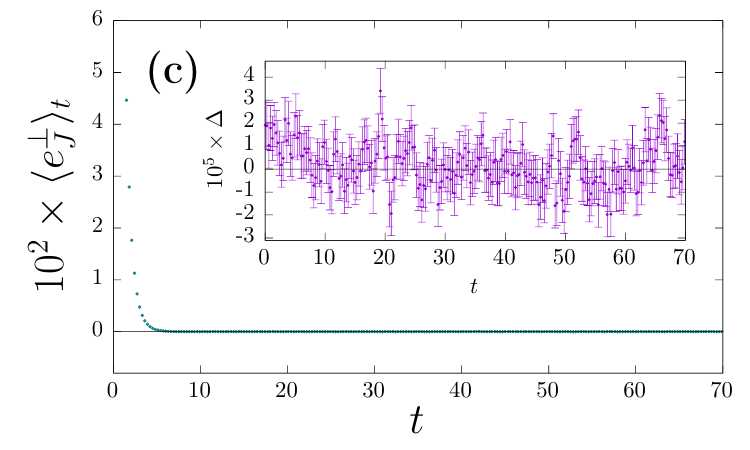}
    \includegraphics[width=0.45\linewidth]{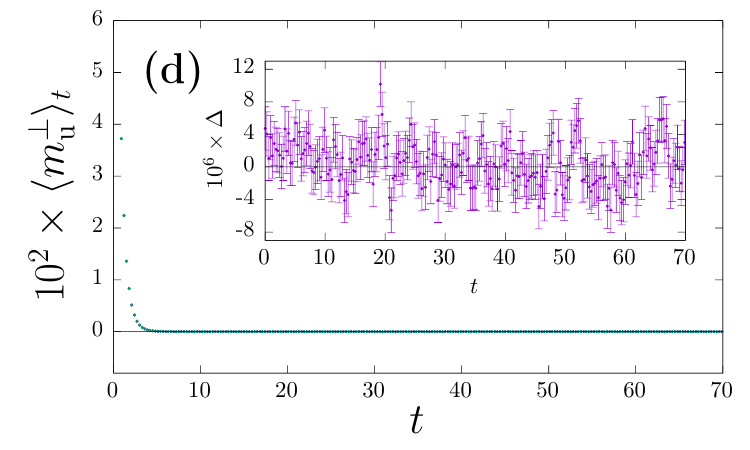}
    \caption{Relaxation to equilibrium of observables for lattice sizes $N = 256\times256$ (purple) and $N= 512\times512$ (green).
                 Note that the green dots occlude the purple ones. The inset shows their difference.
                 (a) Exchange energy per spin as a function of time, 1-temperature jump protocol.
                 (b) Magnetization per spin as a function of time, 1-temperature jump protocol.
                 (c) Exchange energy per spin as a function of time, 2-temperature jump protocol.
                 (d) Magnetization per spin as a function of time, 2-temperature jump protocol.}
     \label{fig:lt_2}
\end{figure}

Let us consider the evolution of one- and two-temperature jump processes of the type
studied in section~\ref{sec:anomalous_relaxation}, wherein the system first undergoes one sudden change
(or two sequential changes) in temperature and then is allowed to relax to thermal equilibrium.
The main panels in figure~\ref{fig:lt_2} show the relaxation of the exchange energy per spin $e_J$
and of the magnetization per spin $m_{\mathrm{u}}$ after subtracting the respective equilibrium values,
\begin{equation}
    \langle{e_J}\rangle_t - e_{J}^\mathrm{eq} = \frac{1}{N}( \langle{\mathcal{E}_{J}}\rangle_t - \mathcal{E}_{J}^\mathrm{eq}) = \langle{{e_J^{\perp}}}\rangle_t,
\end{equation}
\begin{equation}
    \langle{m_{\mathrm{u}}}\rangle_t - m_{\mathrm{u}}^\mathrm{eq}
    =
    \frac{1}{N}( \langle{\mathcal{M}_{\mathrm{u}}}\rangle_t - {\mathcal{M}_{\mathrm{u}}^\mathrm{eq}})
    =
    \langle{{{m}_\mathrm{u}^\perp}}\rangle_t,
\end{equation}
where $\langle \cdot \rangle_t$ denotes the expectation value across simulation trajectories at time $t$,
for the one- and two-jump protocols and lattice sizes of $N=256 \times 256$ and $N=512 \times 512$.
The insets show the corresponding differences, which, again, are smaller than the simulation error.

Similar results are obtained for protocols with fixed $T$ and changes in $h$, or
a mixture thereof.

%%%%%%%%%%%%%%%%%%%%%%%%%%%%%%%%%%%%%%%%%%%%%%%%%%%%%%%%%%%%%%%%%%%%%%%%%%%%%%%%
%%%%%%%%%%%%%%%%%%%%%%%%%%%%%%%%%%%%%%%%%%%%%%%%%%%%%%%%%%%%%%%%%%%%%%%%%%%%%%%%

%%%%%%%%%%%%%%%%%%%%%%%%%%%%%%%%%%%%%%%%%%%%%%%%%%%%%%%%%%%%%%%%%%%%%%%%%%%%%%%%
\section{Effective description of relaxation}\label{sec:effective_description}
%%%%%%%%%%%%%%%%%%%%%%%%%%%%%%%%%%%%%%%%%%%%%%%%%%%%%%%%%%%%%%%%%%%%%%%%%%%%%%%%
Exotic dynamics is typically analyzed in terms of expansions such as Eqs.~\eqref{eq:formal-strong-solution} and~\eqref{eq:A-continuos-time},
where only the leading (and perhaps the first sub-leading) term is kept~\cite{TB26}. This approach is problematic in the thermodynamic limit because,
as we argue in section~\ref{subsect:guess} below, the leading term in the expansion does not have a one-to-one correspondence
to an allegedly leading term in the thermodynamic limit. Rather, the sum in equation~\eqref{eq:A-continuos-time} becomes
an integral over a continuum of time scales.

\begin{figure}[ht]
    \centering
    \includegraphics[width=0.45\linewidth]{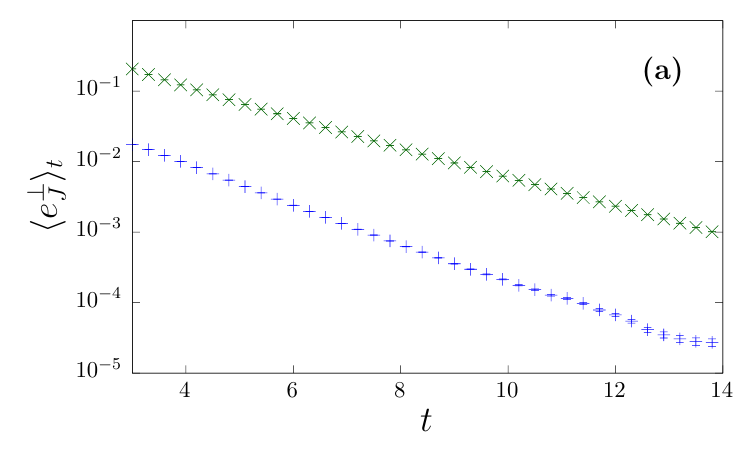}
    \includegraphics[width=0.45\linewidth]{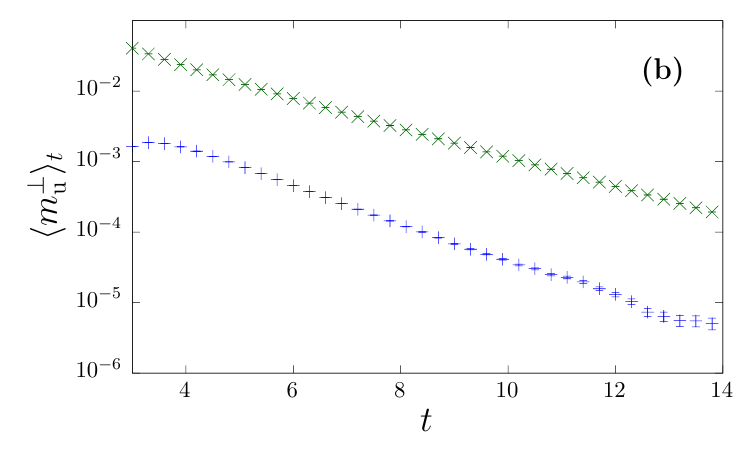}
    \caption{Relaxation curves for (a) $e_J^\perp$ and (b) $m_\text{u}^\perp$ as a function of time in semi-logarithmic scale---data for  $N= 512\times512$ and $T=2.5$ (point E), green for staggered protocol and blue for the random protocol. See section~\ref{sec:effective_description} for the description of both protocols.}\label{fig:presentacion_del_problema}
\end{figure}

To illustrate this problem, in figure~\ref{fig:presentacion_del_problema} we show the relaxation curves for (a) $e_J^\perp$ and (b) $m_\text{u}^\perp$
in a semi-logarithmic plot, where the slopes directly relate to the respective inverse relaxation times $1/\tau$. The time window in the figure corresponds
to the final stages of the relaxation process, up to the time when signal-to-noise ratio decreases below 3.
The figure compares two initial preparations: in the random preparation, spins are independently initialized as $\pm 1$
with 50\% probabilities, while in the staggered preparation spins are initialized as $\sigma_i=(-1)^{x_i+y_i}$.
According to the small-system perspective (see equations~\eqref{eq:formal-strong-solution} and~\eqref{eq:A-continuos-time}),
the time scale $\tau_1$ should be independent of the preparation. Instead, figure~\ref{fig:presentacion_del_problema} shows that
the apparently dominant time scale for the staggered preparation is $\tau_1\approx 2.31$ while for the random preparation is $\tau_1\approx 1.8$.

Nevertheless, with due care, a discrete sum with a common set of time scales can provide an \emph{effective} description of the time evolution
that is accurate-enough for our purposes (section~\ref{sub:analysis_num_data}). Again, this effective description
does not imply that the time scales in the discrete sum have a direct physical interpretation independent of the initial states. 

%%%%%%%%%%%%%%%%%%%%%%%%%%%%%%%%%%%%%%%%%%%%%%%%%%%%%%%%%%%%%%%%%%%%%%%%%%%%%%%%

%%%%%%%%%%%%%%%%%%%%%%%%%%%%%%%%%%%%%%%%%%%%%%%%%%%%%%%%%%%%%%%%%%%%%%%%%%%%%%%%

%%%%%%%%%%%%%%%%%%%%%%%%%%%%%%%%%%%%%%%%%%%%%%%%%%%%%%%%%%%%%%%%%%%%%%%%%%%%%%%%
\subsection{An educated guess for the thermodynamic limit}\label{subsect:guess}
%%%%%%%%%%%%%%%%%%%%%%%%%%%%%%%%%%%%%%%%%%%%%%%%%%%%%%%%%%%%%%%%%%%%%%%%%%%%%%%%
The thermodynamic limit raises some questions about the interpretation of equation~\eqref{eq:A-continuos-time}. 
The numerical results in section~\ref{subsect:TLdynamics} strongly suggest that the time evolution of the \emph{intensive}
version of the observables $E_t[\mathcal{A}^\perp/N]$,
\begin{equation}\label{eq:Tlim-dyn-1}
    E_t[\mathcal{A}^\perp/N]=\sum_{k>1}\, \text{e}^{-|\lambda_k|t/t^{*}}\, \frac{\alpha_k \beta_{k}^{(O)}}{N}\,
\end{equation}
reaches a finite thermodynamic limit for any $t$ (this limit goes to zero as $t\to\infty$).
One may expect the coefficients $\beta_{k}^{(O)}$ to remain of order one in the large-$N$ limit,
\begin{equation}
    \mathbb{E}^{T_b}[O_k]=\langle\mathbf{1}|O_k\rangle=0\,,\quad \mathbb{E}^{T_b}[O_k^2]=\langle O_k|O_k\rangle=1\,,
\end{equation}
because $O_1=\mathbf{1},O_2,O_3,\ldots,O_{2^N}$ are an orthonormal basis of $\mathcal{F}(\Omega)$.
On the other hand, away from a critical point the variance of an intensive quantity is of order $1/N$. Hence,
\begin{equation}\label{eq:Tlim-dyn-2}
   \mathbb[(\mathcal{A}^\perp)^2/N]=\sum_{k>1}\,\frac{\alpha_k^2}{N}\,, 
\end{equation}
should remain of order one in the large-$N$ limit. The simplest way of satisfying simultaneously Eqs.~\eqref{eq:Tlim-dyn-1} and~\eqref{eq:Tlim-dyn-2}
is having only $N$ of the $\alpha_k$ to remain of order one in the large $N$-limit. In particular, having a single coefficient
(say $\alpha_1$) to have a positive large-$N$ limit for $\alpha_1/N$ would imply $\mathbb[(\mathcal{A}^\perp)^2/N]\propto N$.

Hereafter, instead of using the $2^{L\times L}$ eigenvalues $\lambda_i$ of the matrix $R^{\text{HB}}$,
we will describe relaxation process in terms of relaxation times defined by
\begin{equation}
\tau_{i}=-\frac{t^*}{\lambda_{i+1}}\,,\quad i=1,2,3,\ldots,2^{L\times L}-1\,.
\end{equation}
Note that the first eigenvalue $\lambda_1=0$ in equation~\eqref{eq:eigenvalue-ordered}
corresponds to the Boltzmann weight $\pi^{T_b}$ (left eigenvector) or to the constant function $\mathbf{1}$ (right eigenvector).
While one expects $\tau_1$ to remain finite in the $L\to\infty$ limit (because we shall stay away from the critical point),
shorter relaxation times accumulate very quickly towards $\tau=0$. Hence, one should think of an effective density
$\rho_{\mathcal{A}}(\tau)$ of autocorrelation times (most likely a distribution) which is characteristic of the observable
\begin{equation}
\rho_{\mathcal{A}}(\tau)=\lim_{N\to\infty}\sum_{i=1}^{2^{N}-1}\,\delta(\tau-\tau_i)\,\frac{\alpha_{i+1}}{N}\,
\end{equation}
where $\delta$ is Dirac's delta function. Note that although
different observables will have different densities, we expect a common support for the distributions $\rho_{\mathcal{A}}$. 

Then, equation~\eqref{eq:A-continuos-time} takes the form
\begin{equation}\label{eq:evolution-intensive}
\lim_{N\to\infty} E_t[\mathcal{A}^\perp/N]=
\int_0^{\tau_1}\mathrm{d}\tau\, \rho_{\mathcal{A}}(\tau)\,
\text{e}^{-t/\tau}\, \beta^{(O)}(\tau)\,,
\end{equation}
where we expect that the order-one coefficients $\beta_k^{(O)}$ will converge to a  smooth test-function $\beta^{(O)}(\tau)$ in the
large-$N$ limit. 

Let us briefly summarize  the conclusions of our heuristic reasoning: the relaxation to equilibrium
of an intensive observable $\mathcal{A/N}$ is given by  the integral \eqref{eq:evolution-intensive} whose integrand is made of two factors,
the density $\rho_{\mathcal{A}}(\tau)$, which is characteristic of the observable although densities for different observables share a common support,
and the test function $\beta^{(O)}(\tau)$, which defines the preparation of the state.
%%%%%%%%%%%%%%%%%%%%%%%%%%%%%%%%%%%%%%%%%%%%%%%%%%%%%%%%%%%%%%%%%%%%%%%%%%%%%%%%
\subsection{Analysis of numerical data}\label{sub:analysis_num_data}
%%%%%%%%%%%%%%%%%%%%%%%%%%%%%%%%%%%%%%%%%%%%%%%%%%%%%%%%%%%%%%%%%%%%%%%%%%%%%%%%
Unfortunately, in the 2D case even extracting just a few eigenvalues and eigenvectors through exact diagonalizing of the dynamical matrix is unfeasible
for reasonably sized systems. In fact, a similar diagonalization limited to the 4 largest eigenvalues have been recently obtained only for $6\times 6$
lattices with the use of GPUs~\cite{BG24,BG24b}.

Hence, we are limited to the analysis of the numerically obtained $\langle \mathcal{A}^\perp/N\rangle_t$ that would converge to
$E_t[\mathcal{A}^\perp/N]$ only in the limit of an infinite number of simulated trajectories. Extracting the relaxation times $\tau_i$
and the corresponding coefficients from noisy data, which essentially amounts to inverting a Laplace transform, is notoriously difficult~\cite{ES08}.
Therefore, we have turned to fits of our numerical data.

The first alternative that comes to mind (and the one that we have finally followed)
is to fit our numerical data to a linear combination of acertain number $m$ of decreasing exponential functions:
\begin{equation}\label{eq:ansatz-1}
    \langle \mathcal{A}^\perp/N\rangle_t=\sum_{j=1}^m\,a_j\text{e}^{-t/\hat{\tau}_j}\,\quad \text{ for }\quad t \in [t_{\text{min}},t_\text{max}]\,.
\end{equation}
The determination of $t_\text{max}$ does not present particular difficulties: we have stopped at the shortest time such that
$\langle \mathcal{A}^\perp/N\rangle$ is smaller than three times its simulation error. On the other hand, the accuracy and stability of the
parameters $a_j$ and $\hat{\tau}_j$ depend strongly on the choices for $m$ and $t_\text{min}$. 

An alternative, which may seem  more natural for a continuous distribution $\rho_\mathcal{A}(\tau)$ (see~equation~\eqref{eq:evolution-intensive})
would be
\begin{equation}\label{eq:ansatz-2}
    \langle \mathcal{A}^\perp/N\rangle_t=\frac{\text{e}^{-t/\hat{\tau}^*}}{t^{b_0}}\sum_{k=0}^{m^*}\frac{c_k}{t^k}\,\quad \text{ for }\quad t \in [t_{\text{min}},t_\text{max}]\,.
\end{equation}

\begin{figure}[t]
    \centering
\includegraphics[width=0.32\linewidth]{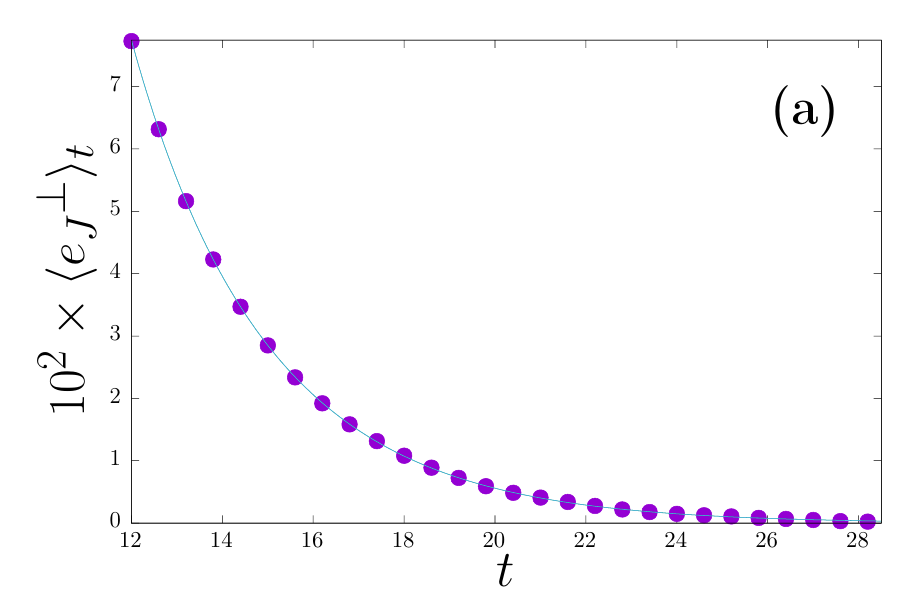}
\includegraphics[width=0.32\linewidth]{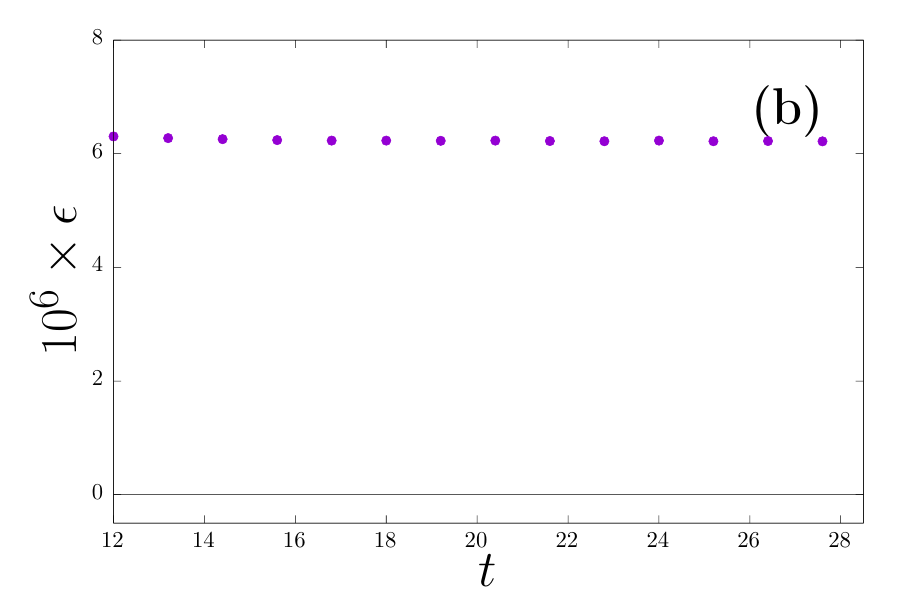}
 \includegraphics[width=0.32\linewidth]{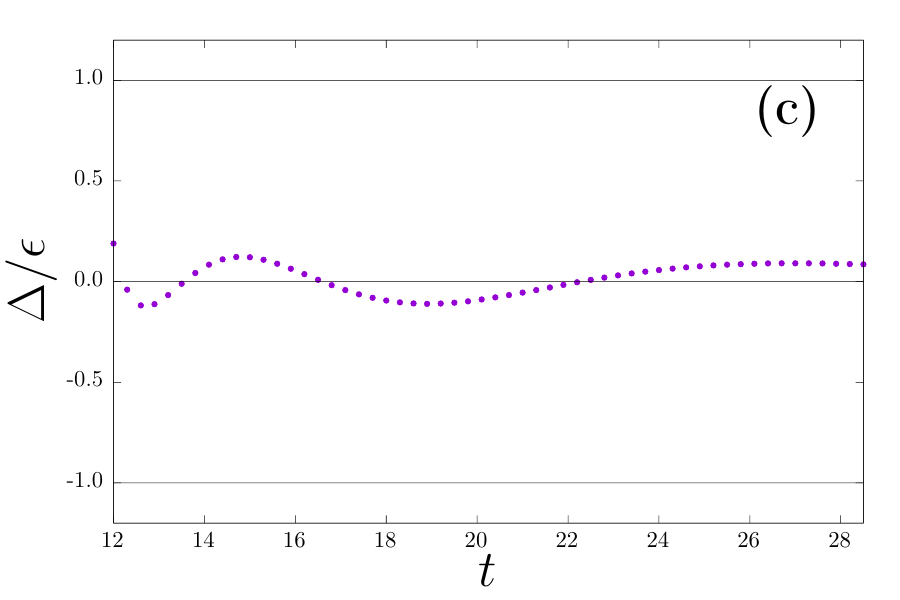}
\caption{Fits to simulation data using the two functional forms (\ref{eq:ansatz-1}) and (\ref{eq:ansatz-2})
              for ${\langle{e_{J}^{\perp}}\rangle}_t$ at point B ($T=0.067, h=4.01$), obtained over the fitting window 
             [$t_\mathrm{min}$, $t_\mathrm{max}$].
             (a) Comparison of fits (blue and green curves) to data (purple dots).
             The dots are bigger than the error to aid visualization.
             (b) Simulation error as a function of time.
             (c) Difference between the two ansatzse normalized by the simulation error.}\label{fig:ans_combined}
\end{figure}

Deciding among the two functional forms is largely a matter of taste. In fact, the $\chi^2$ test is of very limited value given the strong statistical
correlation for $\langle \mathcal{A}^\perp/N\rangle_t$ at different times. For simplicity, we have computed the $\chi^2$ using only the diagonal
terms of the covariance matrix, because getting the full covariance matrix with enough accuracy to allow for a safe matrix inversion
is computationally very expensive. As a consequence, we have found values of $\chi^2$ significantly smaller than the nominal number of degrees of freedom.
Finding $p$-values larger than 0.99 (with both fitting ansatzes) has been the rule rather than the exception. Indeed, not only do our fits to
equation~\eqref{eq:ansatz-1} and to equation~\eqref{eq:ansatz-2} represent the data exceedingly well, as shown in figure~\ref{fig:ans_combined},
but the difference between the two fits is about one-order of magnitude smaller than the simulation error. We have decided
to use equation~\eqref{eq:ansatz-1} for two reasons: simplicity, and the more direct formal similarity between the coefficients $a_k$
and the product of coefficients $\alpha_k\beta^{(O)}_k$, and between the effective time scales $\hat{\tau}_k$  and the microscopic time scales $\tau_k$.

Our strategy has been as follows. We work with $\mathcal{A}=\mathcal{M}$ and use as starting point for the simulation an antiferromagnetic initial state
$\sigma_i=(-1)^{x_i+y_i}$, which is farther from equilibrium than both the uniform and the random initial states. As we said earlier, we have determined
$t_\text{max}$ as the shortest time such that (the absolute value of) $\langle \mathcal{A}^\perp/N\rangle_t$ becomes smaller than three times its statistical error,
thus avoiding too small signal-to-noise ratio.  Our next step is setting $m=1$, and fixing $t_\text{min}$ as the smallest time for which we get a $p$-value above 0.5.
Then, keeping fixed $a_1$ and $\hat{\tau}_1$, we set $m=2$ and decrease $t_\text{min}$ following a similar criterion, thus finding $a_2$ and $\hat{\tau}_2$.
Then, taking these values as initial values and for the same $t_\text{min}$, we carry out a new fit in which $a_1,a_2,\hat{\tau}_1$, and $\hat{\tau}_2$
are free to vary. The process of increasing $m$ and decreasing $t_\text{min}$ is iterated until a maximum value $m=4$.
As a consistency check, documented in Table~\ref{tab:compat_1}, the same set of effective timescales $\hat{\tau}_k$ accounts for relaxation processes
of other observables and other starting points (in these fits the only free parameters are the amplitudes). 

When analyzing the more complex dynamical processes considered in section~\ref{sec:anomalous_relaxation}, we have taken the
set of effective timescales $\hat{\tau}_k$ appropriate for the final bath temperature. Hence, the behavior of the different protocols
will be interpreted in terms of their respective amplitudes $a_k$, as obtained from fits with the same time window $[t_\text{min},t_\text{max}]$.

Finally, we mention that the high correlation of the data makes the error estimation using only the diagonal elements of the correlation matrix typically
used with the Levenberg-Marquart fitting algorithm overly optimistic. To obtain a more realistic error estimate, we have resorted to the jackknife algorithm,
which is standard in the field of Monte Carlo simulations~\cite{AM:05}, which, in the context of fits to correlated data, is described in detail in
reference~\cite{YL11}.

Our simulations consist of $M$ time trajectories for each observables $A^\perp(t)/N$.  Time is sampled at  $t_n = n \Delta t$ $(n=1,2,3,\ldots)$,
and we ensure that our trajectories are statistically independent. We split the $M$ trajectories into 200 disjoint sets of trajectories,
termed blocks (the number of trajectories is the same in all blocks). Then, we obtain the averaged trajectory for the $l$-th block,
$A_l^\perp(t)$, and the $r$-th jackknife block $A_{r,jk}^\perp(t)$ is obtained by averaging $A_l^\perp(t)$ over $l\neq r$.
A fit is then carried out for all the $A_{r,jk}^\perp(t)$, using only the diagonal part of the covariance matrix
(in other words, $A_{r,jk}^\perp(t_1)$ and $A_{r,jk}^\perp(t_2)$ are treated as independent stochastic variables if $t_1\neq t_2$).
The (diagonal) covariance matrix in all fits is the one obtained from the full set of $M$ trajectories.
Finally, the errors in the fit parameters are obtained from the fluctuations of the fit parameters as obtained from the fits for the 200 jackknife blocks.

%%%%%%%%%%%%%%%%%%%%%%%%%%%%%%%%%%%%%%%%%%%%%%%%%%%%%%%%%%%%%%%%%%%%%%%%%%%%%%%%
%%%%%%%%%%%%%%%%%%%%%%%%%%%%%%%%%%%%%%%%%%%%%%%%%%%%%%%%%%%%%%%%%%%%%%%%%%%%%%%%

%%%%%%%%%%%%%%%%%%%%%%%%%%%%%%%%%%%%%%%%%%%%%%%%%%%%%%%%%%%%%%%%%%%%%%%%%%%%%%%%
\begin{figure}[t]
    \centering
\includegraphics[width=0.32\linewidth]{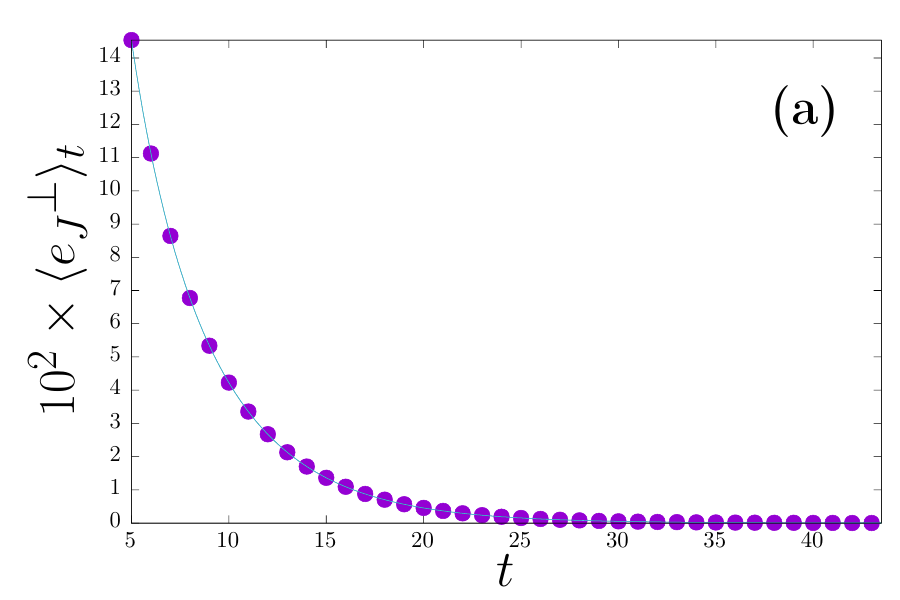}
\includegraphics[width=0.32\linewidth]{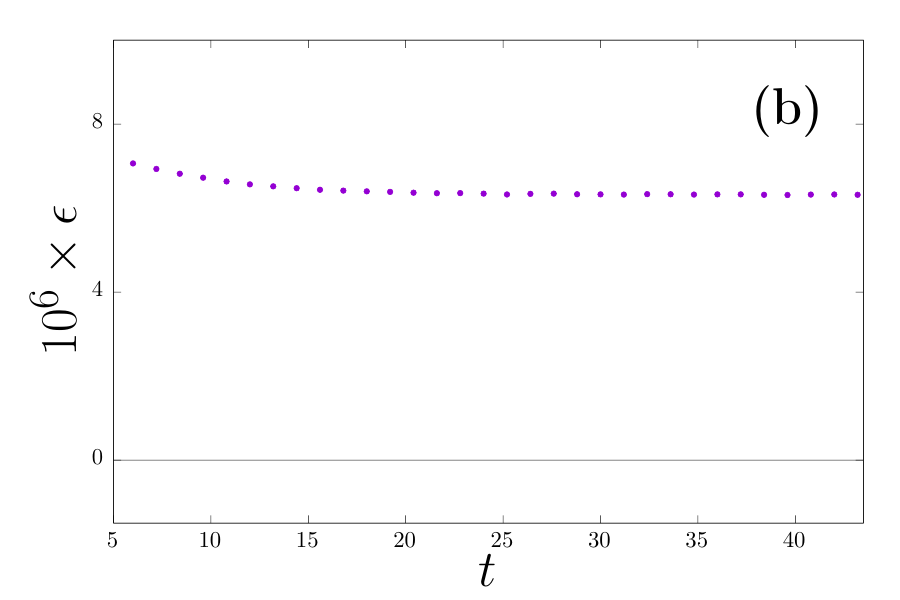}
 \includegraphics[width=0.32\linewidth]{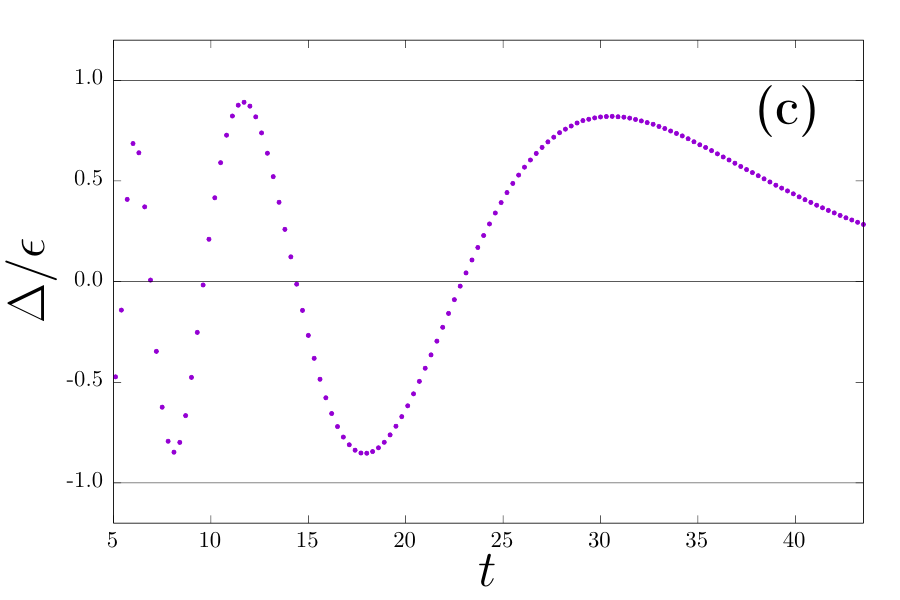}
\caption{Fits to simulation data using the two functional forms (\ref{eq:ansatz-1}) and (\ref{eq:ansatz-2}) for ${\langle{e_J^{\perp}}\rangle}_t$
              at point C ($T=0.5, h=4.01$), obtained over the fitting window  [$t_\mathrm{min}$, $t_\mathrm{max}$].
              (a) Comparison of fits (blue and green curves) to data (purple dots). The dots are bigger than the error to aid visualization.
              (b) Simulation error as a function of time.
              (c) Difference between the two ansatzes normalized by the simulation error.}\label{fig:ans_combined_2}
\end{figure}
%%%%%%%%%%%%%%%%%%%%%%%%%%%%%%%%%%%%%%%%%%%%%%%%%%%%%%%%%%%%%%%%%%%%%%%%%%%%%%%%

%%%%%%%%%%%%%%%%%%%%%%%%%%%%%%%%%%%%%%%%%%%%%%%%%%%%%%%%%%%%%%%%%%%%%%%%%%%%%%%%
\begin{table}
\caption{Goodness-of-fit results for the ansatz ~\ref{eq:ansatz-1} for both magnetization and exchange energy, using $m = 2$
              exponentials with timescales $\hat{\tau}_{1}$ and $\hat{\tau}_{2}$ determined for each of the points in table~\ref{tab:working_points}
              over the time window $[t_{\mathrm{min}}, t_{\mathrm{max}}]$, and for each one of starting configurations of the spin field:
              R (Random), U (Uniform) and S (Staggered), as described in section~\ref{sub:analysis_num_data}.
              Reported values include $\chi^{2}/\text{dof}$ (computed using only the diagonal terms of the covariance matrix) and the corresponding
              $p$-values to five decimal places to illustrate the high correlation between the data, although this precision is unnecessarily high
              for the reported values of $\chi^{2}/\text{dof}$.}
\centering
\begin{tabular}{ccc c rr r@{/}lc r@{/}lc}
\hline
 Point & $\hat{\tau}_{1}$ & $\hat{\tau}_{2}$ & Start & $t_{\mathrm{min}}$ & $t_{\mathrm{max}}$ & $\chi^{2}$ &
              dof ($m$) & $p$-value ($m$) & $\chi^{2}$ & dof ($e_{J}$) & $p$-value ($e_{J}$)\\
\hline
A & 0.9350 & 0.4271 & R & 5.8 & 12.0 & 77.6 & 123 & 0.99953 & 80.9 & 123 & 0.99877 \\
 & & & U & 2.2 & 8.8 & 81.0 & 131 & 0.99982 & 81.0 & 131 & 0.99982  \\
 & & & S & 6.0 & 12.3 & 86.4 & 125 & 0.99661 & 86.4 & 125 & 0.99661 \\
B & 2.5462 & 2.11 & R & 10.0 & 17.4 & 104.2 & 147 & 0.99700 & 104.2 & 147 & 0.99700 \\
 & & & U & 10.7 & 17.3 & 85.7 & 131 & 0.99923 & 85.7 & 131 & 0.99923 \\
 & & & S & 21.0 & 29.0 & 98.1 & 159 & 0.99996 & 98.1 & 159 & 0.99996 \\
C & 4.8059 & 2.3926 & R & 16.0 & 27.3 & 163.9 & 225 & 0.99921 & 162.4 & 125 & 0.99942 \\
 &  &  & U & 12.0 & 25.0 & 174.4 & 259 & 0.99999 & 173.7 & 259 & 0.99999 \\
 &  &  & S & 22.0 & 43.5 & 357.7 & 429 & 0.99479 & 359.6 & 429 & 0.99358 \\
D & 12.7251 & 8.8145 & R & 38.0 & 61.0 & 102.9 & 151 & 0.99902 & 102.8 & 151 & 0.99904 \\
 & & & U & 35.0 & 61.8 & 135.8 & 177 & 0.99072 & 136.2 & 177 & 0.99007 \\
 & & & S & 80.0 & 107.0 & 127.4 & 178 & 0.99841 & 126.9 & 178 & 0.99858 \\
E & 2.3111 & 1.6324 & R & 9.0 & 14.2 & 73.1 & 103 & 0.98861 & 64.7 & 103 & 0.99884 \\
 & & & U & 9.0 & 14.0 & 78.9 & 99 & 0.93167 & 63.2 & 99 & 0.99806 \\
 & & & S & 16.0 & 22.4 & 74.5 & 127 & 0.99994 & 88.4 & 127 & 0.99633 \\
\hline
\end{tabular}
\label{tab:compat_1}
\end{table}
%%%%%%%%%%%%%%%%%%%%%%%%%%%%%%%%%%%%%%%%%%%%%%%%%%%%%%%%%%%%%%%%%%%%%%%%%%%%%%%%

%%%%%%%%%%%%%%%%%%%%%%%%%%%%%%%%%%%%%%%%%%%%%%%%%%%%%%%%%%%%%%%%%%%%%%%%%%%%%%%%
\section{Anomalous relaxation processes}\label{sec:anomalous_relaxation}
%%%%%%%%%%%%%%%%%%%%%%%%%%%%%%%%%%%%%%%%%%%%%%%%%%%%%%%%%%%%%%%%%%%%%%%%%%%%%%%%

%%%%%%%%%%%%%%%%%%%%%%%%%%%%%%%%%%%%%%%%%%%%%%%%%%%%%%%%%%%%%%%%%%%%%%%%%%%%%%%%
\begin{figure}[ht]
    \centering
    \includegraphics[width=0.48\linewidth]{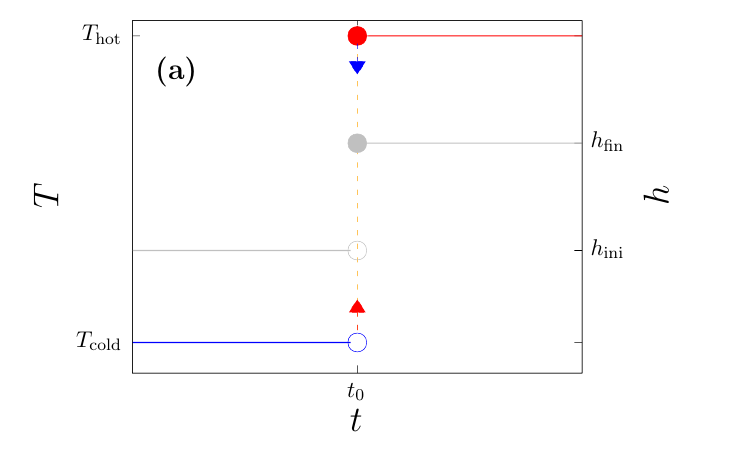}
    \hfill
    \includegraphics[width=0.48\linewidth]{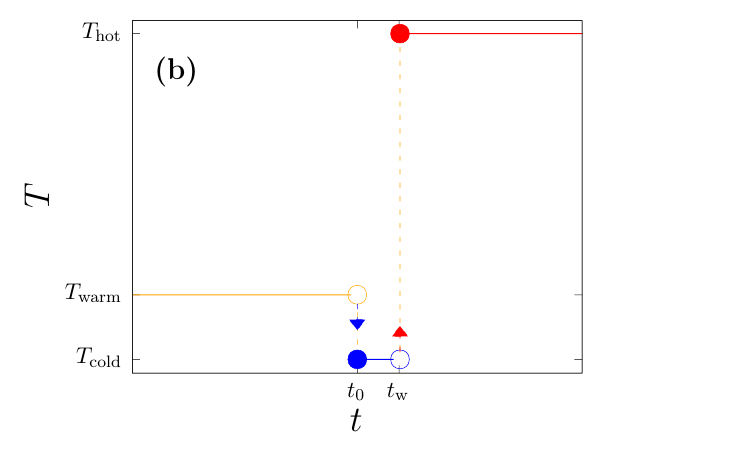}
    \medskip
    \includegraphics[width=0.48\linewidth]{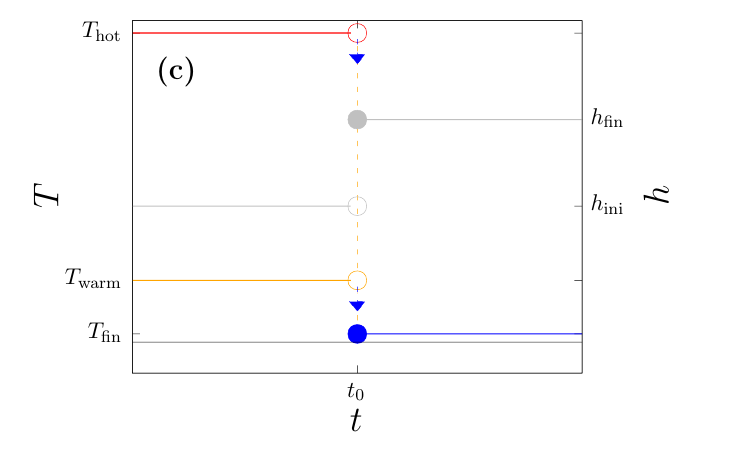}
    \hfill
    \includegraphics[width=0.48\linewidth]{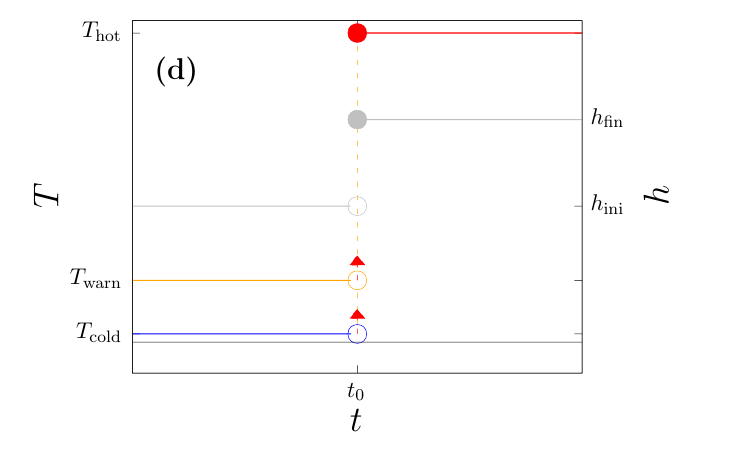}
    \caption{Sketch of the time evolution of the bath temperature for our different types of protocols.
    (a) One-jump temperature protocol $\text{P} \to Q$, which might include a jump in $h$.
    (b) Two-jump temperature protocol $\text{P} \to Q \to \text{R}$. Note that our two-jump protocols are performed at constant field strength,
    so $h$ is not shown in this panel.
    (c) Direct Mpemba protocol, which might include a jump in $h$.
    (d) Inverse Mpemba protocol, which might include a jump in $h$.}
    \label{fig:N-jump}
\end{figure}
%%%%%%%%%%%%%%%%%%%%%%%%%%%%%%%%%%%%%%%%%%%%%%%%%%%%%%%%%%%%%%%%%%%%%%%%%%%%%%%%

In this section  we demonstrate how certain anomalous relaxation effects (asymmetric heating/cooling, precooling/heating protocols,
and both direct and inverse Mpemba effects) can be not only reproduced in our 2D setting, but predicted using Figs.~\ref{fig:crit-suscept}
and~\ref{fig:N-jump}, to which we make constant reference along this section. We encourage the reader to have both figures at hand to follow the discussion.
We stress our careful handling of the thermodynamic limit and the fact that we deal with two different parameter changes:
the usual one, which is the temperature $T$, and a second one, the magnetic field $h$.

Our main hypothesis is an adaptation to the thermodynamic limit of the main finding in reference~\cite{GL24}.
In the notation of equation~\eqref{eq:evolution-intensive}, our ansatz reads
\begin{equation}\label{eq:our-main-hyp}
    \int_{\tau_1-\Delta \tau}^{\tau_1}\mathrm{d}\tau\, \rho_{\mathcal{A}}(\tau)\, \beta^{(O)}(\tau)\,
    \approx
    \frac{1}{\Lambda_{\mathcal{A}}} \left( E_{\tw}[\mathcal{M}^2_{\mathrm{st}}] -  \mathbb{E}^{T_{\text{f}}}[\mathcal{M}^2_{\mathrm{st}}]\right)\,,
\end{equation}
where $T_{\text{f}}$ is the final bath temperature, $\Lambda_{\mathcal{A}}$ is a constant independent of the preparation protocol,
and $\tw$ is the time elapsed before the bath temperature is  set to $T_{\text{f}}$.

For instance, in the simplest protocol the bath temperature is suddenly changed from $T_{\text{i}}$ to $T_{\text{f}}$
and therefore $E_{\tw}[\ldots]=\mathbb{E}^{T_{\text{i}}}[\ldots]$ provided that system is already in equilibrium at time $\tw$.
There are two important remarks to be made. First, the loosely defined time interval $\tau_1-\Delta \tau < \tau< \tau_1$
should include all the time scales relevant  for the relaxation process. Second, the left-hand side of equation~\eqref{eq:our-main-hyp}
explicitly depends on the observable $\mathcal{A}$, while the only dependence on $\mathcal{A}$ of the right-hand side is through
the global prefactor $\Lambda_{\mathcal{A}}$. Our ansatz~\eqref{eq:our-main-hyp} is physically based on Langer's theory for the
nucleation time~\cite{LA67,PA88}, which suggests that dominant time scales are induced by the fluctuations of the metastable
phase (recall that we shall be working in the part of the phase diagram where the antiferromagnetic phase is metastable,
labeled as AFM in  figure~\ref{fig:crit-suscept}(a); see also equation~\eqref{eq:tau_xist_dyn_exp}).

We will see how the ansatz~\eqref{eq:our-main-hyp} in conjunction with figure~\ref{fig:crit-suscept}(b) allows us
to predict the evolution of the system. A first crucial consequence of our ansatz is that if the right-hand side of equation~\eqref{eq:our-main-hyp}
vanishes for a certain preparation protocol, it vanishes for any observable $\mathcal{A}$. Therefore, that protocol results in an
exponential speed-up independently of the particular observable under consideration. This is why, standing in contrast to other works,
our anomalous dynamic effects occur simultaneously in two physical quantities: the exchange energy and the magnetization. 

In practice, we cannot use the left-hand side of equation~\eqref{eq:our-main-hyp} as such, and we will replace it with
the dominant amplitudes $a_j$ obtained from the fitting process described in section~\ref{sub:analysis_num_data}.
Also, when assessing the numerical results in this section, keep in mind that the equilibrium expectation value
for the fluctuating part $\mathcal{A}^\perp$ is zero.

Throughout this section, we shall use the following notation:
\begin{itemize}
\item One-jump protocols, see figure \ref{fig:N-jump}(a), (c) and (d), will be denoted as  $\text{P}\to\text{Q}$.
         In all cases the system starts at an initial equilibrium state corresponding to $\text{P}=(h_{\mathrm{ini}}, T_{\mathrm{fin}})$.
         Then,  either $h$, $T$ or both are suddenly changed to the coordinates of the second point $\text{Q}=(h_{\mathrm{ini}}, T_{\mathrm{fin}})$,
         where the system is allowed to relax towards the new equilibrium.
\item Two-jump processes, see figure \ref{fig:N-jump}(b), will be denoted as $\text{P} \to \text{R} \to \text{Q}$.
         The system starts at the initial equilibrium state corresponding to $\text{P}=(h_{\mathrm{ini}}, T_{\mathrm{ini}})$.
         Next, either $h$, $T$ or both are suddenly changed to the values corresponding to the intermediate point
         $\text{R}=(h_{\mathrm{int}}, T_{\mathrm{int}})$ and the system is allowed to relax for some time $\tw$ much less than the (global) relaxation
         time at $\text{R}$. In the final step, $h$, $T$ or both are suddenly changed  again to the final point in the phase diagram
        $\text{Q}=(h_{\mathrm{fin}}, T_{\mathrm{fin}})$, where the system is allowed to relax to equilibrium. 
\end{itemize}

We stress again that our two-dimensional phase diagram permits large changes in correlation length and consequently
in equilibration times (see equation~\eqref{eq:tau_xist_dyn_exp}) that could not be achieved in previous works.
%%%%%%%%%%%%%%%%%%%%%%%%%%%%%%%%%%%%%%%%%%%%%%%%%%%%%%%%%%%%%%%%%%%%%%%%%%%%%%%%
\subsection{Asymmetric heating/cooling processes}\label{sub:asymmetric}
%%%%%%%%%%%%%%%%%%%%%%%%%%%%%%%%%%%%%%%%%%%%%%%%%%%%%%%%%%%%%%%%%%%%%%%%%%%%%%%%
In this section we discuss one-jump precesses where the forward process $\text{P}\to \text{Q}$ and the backward process $\text{Q}\to \text{P}$
show a marked difference in the time required to reach equilibrium. A recurring theme will be that by looking at figure~\ref{fig:crit-suscept}
we can predict the qualitative results of our simulations: since the staggered susceptibility
scales as $\xi_{\text{st}}^{7/4}$ and $\tau_1$ scales as $\xi_{\text{st}}^z$ ($z\approx 2.16$, see equation~\eqref{eq:tau_xist_dyn_exp}),
it follows that if $\text{P}$ and $\text{Q}$ differ significantly in $\chi_{\text{st}}$, the corresponding relaxation times will also differ significantly,
and the relaxation towards the point having the largest $\chi_{\text{st}}$ will be the slower.

We will compare a process in which only the temperature is varied with a process in which both the temperature and the magnetic
field are varied, and show that even in the simpler case the faster process can be either the cooling or the heating one.

\begin{figure}[ht]
    \centering
    \includegraphics[width=0.48\linewidth]{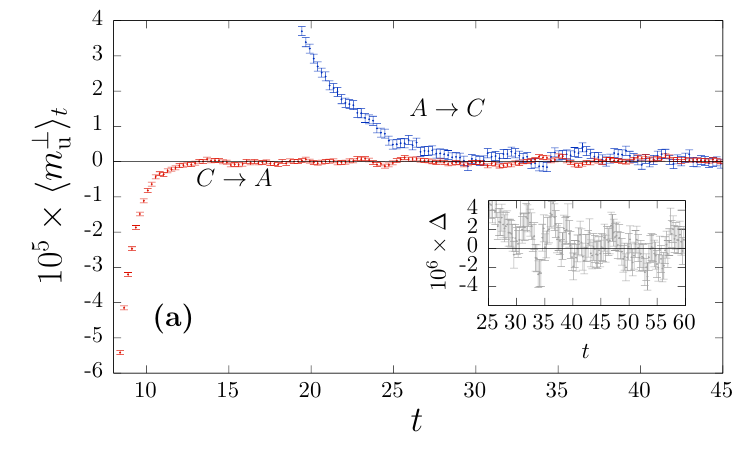}
    \hfill
    \includegraphics[width=0.48\linewidth]{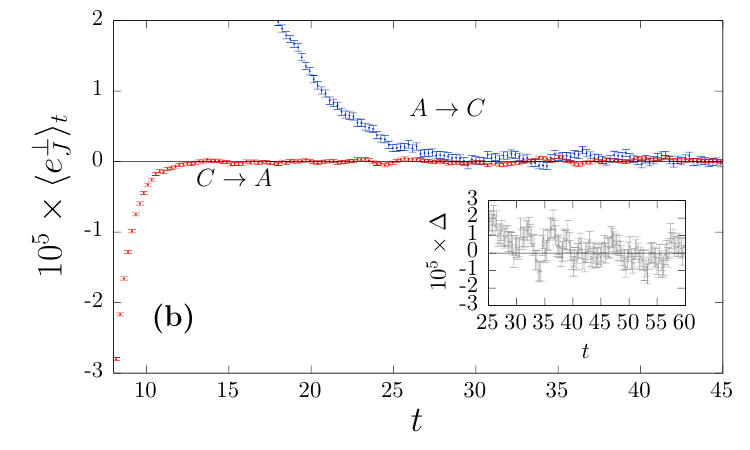}
    \includegraphics[width=0.48\linewidth]{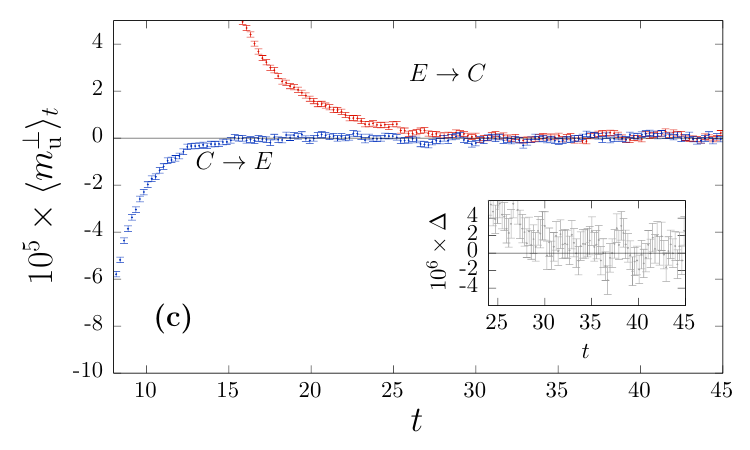}
    \hfill
    \includegraphics[width=0.48\linewidth]{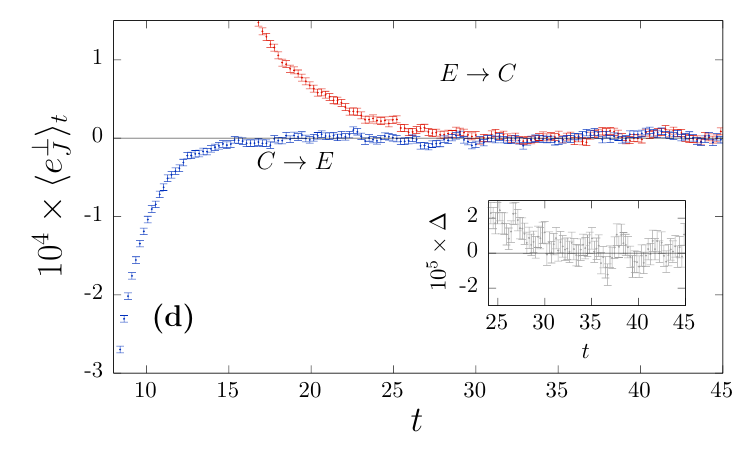}
    \vspace{0.5em}
    \caption{Asymmetric relaxation to equilibrium between points with a constant value of $h=4.01$ of observables:
    (a) magnetization, between $\text{A}$ and $\text{C}$;
    (b) exchange energy, between $\text{A}$ and $\text{C}$;
    (c) magnetization, between $\text{E}$ and $\text{C}$, and
    (d) exchange energy, between $\text{E}$ and $\text{C}$.
    Red curves come from hotter initial temperatures (cooling processes),
    while blue curves come from colder ones (heating processes).
    Insets show the difference between each pair of curves.}
    \label{fig:asymmetry-sameh}
\end{figure}

Consider first the heating process $\text{A}\to\text{C}$ and its reversed cooling process $\text{C}\to\text{A}$, which correspond
to protocols (a) and (c)  in figure~\ref{fig:N-jump} respectively. Note that the value of $h$ is fixed for both processes.
The locations of $\text{A}$ and $\text{C}$ are marked in the phase diagram in figure~\ref{fig:crit-suscept}(a),
and the corresponding susceptibilities can be seen in figure~\ref{fig:crit-suscept}(b).
Since $\chi_{\text{st}}(\text{C})\approx 50 \chi_{\text{st}}(\text{A})$, we expect the heating process $\text{A}\to \text{C}$ to be slower
than the cooling process $\text{C}\to \text{A}$. In figure~\ref{fig:asymmetry-sameh}(a) and (b) we compare the relaxation to equilibrium
for both processes and for two observables, (a)  $\langle {m_{\mathrm{u}}^{\perp}}(t) \rangle$ and (b) $\langle {e_J^{\perp}}(t) \rangle$. 
As we anticipated, the time to reach equilibrium for both observables is larger for the heating process $\text{A}\to \text{C}$ by a factor of three.
Although this result matches our qualitative expectation, quantitative scaling arguments predict a much larger ratio for the time to equilibrium
$(\xi_{\text{st}}(\text{C})/\xi_{\text{st}}(\text{A}))^z\approx (1.29/0.11)^{2.16}\approx 204$, see table~\ref{tab:xist}.
This discrepancy is not surprising because both $\text{A}$ and $\text{C}$ are far away from the region of the phase diagram where the scaling
relations are accurate (the smallest of the two correlation lengths is only 0.11 lattice spacings).

The heating process $\text{C}\to\text{E}$ (blue curves) and the cooling process $\text{E}\to\text{C}$ (red curves)
in figure~\ref{fig:asymmetry-sameh} (c) and (f) differ from their analogs in the upper panels of figure~\ref{fig:asymmetry-sameh}
in that the slower of the two processes is the cooling one. Indeed, of the two points in the phase diagram,
it is $\text{C}$ that has the larger $\chi_\text{st}$ (and hence the larger $\xi_{\text{st}}$) in figure~\ref{fig:crit-suscept}(b).
Our simple scaling argument predicts that the time to equilibrium for both processes will be in the ratio
$(\xi_{\text{st}}(\text{C})/\xi_{\text{st}}(\text{E}))^z\approx (1.29/0.91)^{2.16}\approx 2.12$. The data in figure~\ref{fig:asymmetry-sameh}(c)
and (f) display a time to equilibrium of approximately 18 for $\text{C}\to\text{E}$ and of 28 for $\text{E}\to\text{\text{C}}$, with a ratio
$28/18=1.55...$. As we see, the prediction of the scaling argument is getting closer to the real data as the values $\xi_{\text{st}}$ increase.

%%%%%%%%%%%%%%%%%%%%%%%%%%%%%%%%%%%%%%%%%%%%%%%%%%%%%%%%%%%%%%%%%%%%%%%%%%%%%%%%
%%%%%%%%%%%%%%%%%%%%%%%%%%%%%%%%%%%%%%%%%%%%%%%%%%%%%%%%%%%%%%%%%%%%%%%%%%%%%%%%

Next we discuss processes involving point D, which has a different value of the magnetic field ($h=3.9$ rather than $h=4.01$).
This change in the magnetic field entails a larger correlation length $\xi_{\text{st}}$ and, as we will see, a more accurate estimate
of the order of magnitude for the time asymmetry from the scaling relation. 

\begin{figure}[ht]
    \centering
    \includegraphics[width=0.48\linewidth]{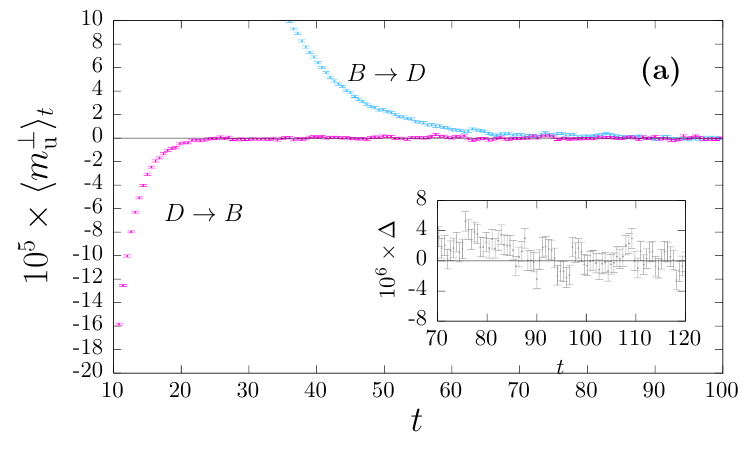}
    \hfill
    \includegraphics[width=0.48\linewidth]{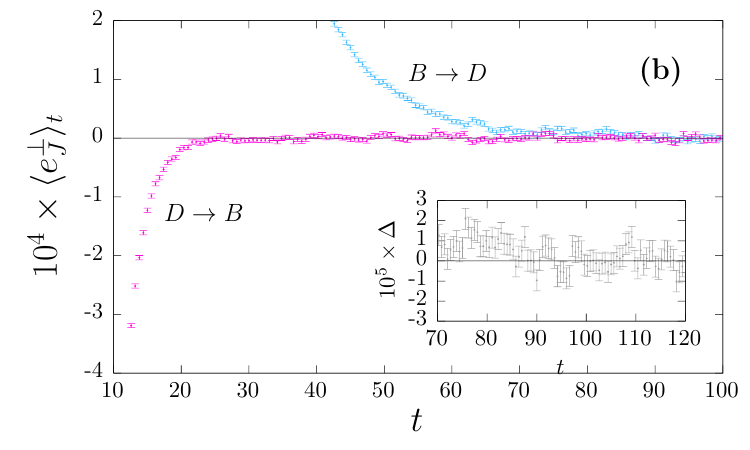}
    \vspace{0.5em}
    \includegraphics[width=0.48\linewidth]{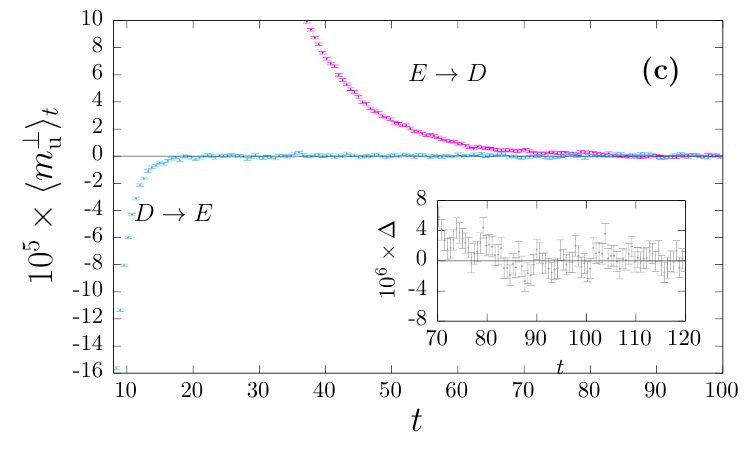}
    \hfill    
    \includegraphics[width=0.48\linewidth]{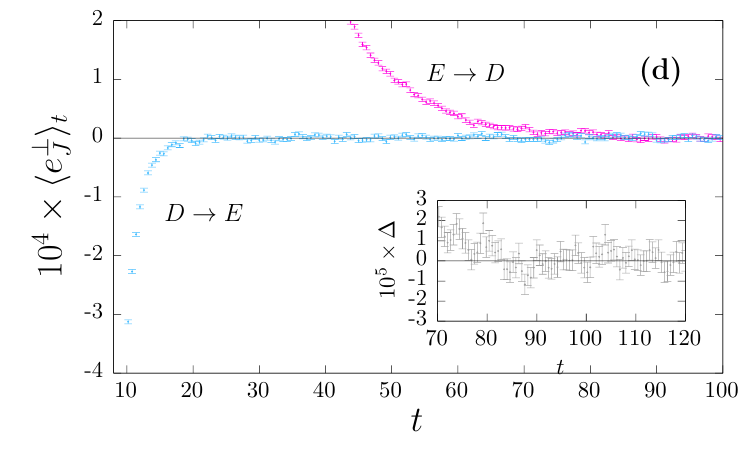}
    \caption{Asymmetric relaxation to equilibrium between points with a jump in the value of $h$ for observables:
    (a) magnetization, between $\text{B}$ $(h=4.01)$ and $\text{D}$ $(h=3.9)$;
    (b) exchange energy, between $\text{B}$ $(h=4.01)$ and $\text{D}$ $(h=3.9)$;
    (c) magnetization, between $\text{D}$ $(h=3.9)$ and $\text{E}$ $(h=4.01)$; and
    (d) exchange energy, between $\text{D}$ $(h=3.9)$ and $\text{E}$ $(h=4.01)$.
    Magenta curves come from hotter initial temperatures (cooling processes),
    while cyan curves come from colder ones (heating processes). Insets show the difference between each pair of curves.}
    \label{fig:asymmetry-jumph}
\end{figure}

Let us first consider the heating process between states with different values of the magnetic field $\text{B}\to\text{D}$
and its reverse cooling process $\text{D}\to\text{B}$. The by now familiar argument reads as follows:
since $\chi_{\text{st}}(\text{D})>\chi_{\text{st}}(\text{B})$ (see figure~\ref{fig:crit-suscept}) and hence $\xi_{\text{st}}(\text{D})>\xi_{\text{st}}(\text{B})$,
the process ending in D (heating) must be the slowest by a factor close to (see table~\ref{tab:xist}) 
$(\xi_{\text{st}}(\text{D})/\xi_{\text{st}}(\text{B}))^z\approx (1.99/1.05)^{2.16}\approx 3.97\ldots$.
One may compare this estimate with the data in figure~\ref{fig:asymmetry-jumph}(a) and (b).
The time to equilibrium is consistent for both observables $m_\text{u}$ and $e_J$. In the case of the cooling process
$\text{D}\to \text{B}$ the time to equilibrium around 20. For the heating process it is more difficult to give a precise value:
any value between 70 and 90 could be acceptable. Taking 80 as compromise, we get a ratio $80/20=4$, unreasonably
close to the ratio predicted by the scaling relation (keep in mind that the largest correlation length involved in the process
is still smaller than 2 lattice spacings).

Finally, we have considered the heating $\text{D}\to \text{E}$ and reverse cooling $\text{E}\to \text{D}$ processes.
Again, $\chi_{\text{st}}(\text{D})>\chi_{\text{st}}(\text{E})$, hence $\xi_{\text{st}}(\text{D})>\xi_{\text{st}}(\text{E})$,
which suggests that the slowest process will be the one ending in D by a factor close to
$(\xi_{\text{st}}(\text{D})/\xi_{\text{st}}(\text{E}))^z\approx (1.99/0.91)^{2.16}= 5.42\ldots$.
The data in figure~\ref{fig:asymmetry-jumph}(c) and (d), shows consistent times to equilibrium for both observables.
For the heating process $\text{D}\to \text{E}$ the equilibration time is around 18.
For the cooling process $\text{E}\to \text{D}$ any value between 80 and 90 is acceptable.
Taking 85 time units as a compromise, we get a ratio of 85/18 = 4.72\ldots, which is reasonably
close to the prediction of the  scaling relation.

%%%%%%%%%%%%%%%%%%%%%%%%%%%%%%%%%%%%%%%%%%%%%%%%%%%%%%%%%%%%%%%%%%%%%%%%%%%%%%%%
%%%%%%%%%%%%%%%%%%%%%%%%%%%%%%%%%%%%%%%%%%%%%%%%%%%%%%%%%%%%%%%%%%%%%%%%%%%%%%%%

%%%%%%%%%%%%%%%%%%%%%%%%%%%%%%%%%%%%%%%%%%%%%%%%%%%%%%%%%%%%%%%%%%%%%%%%%%%%%%%%
\subsection{Precooling protocols}\label{sub:precooling}
%%%%%%%%%%%%%%%%%%%%%%%%%%%%%%%%%%%%%%%%%%%%%%%%%%%%%%%%%%%%%%%%%%%%%%%%%%%%%%%%
Precooling protocols are two-jump protocols intended to achieve a faster approach to equilibrium at a higher temperature
by a short excursion to a \emph{lower} temperature for a system initially at equilibrium at an intermediate temperature.
The key idea can be found in the right-hand side of equation~\eqref{eq:our-main-hyp}: if the time duration of the low-temperature
excursion $\tw$ is such that the expected value of $\mathcal{M}_{\text{st}}^2$ is near to the equilibrium value at the final temperature,
the left-hand side of equation~\eqref{eq:our-main-hyp} will be suppressed. This implies that the contribution
of the large relaxation times $\tau$ to the equilibrium at the final temperature will be anomalously small
as compared to the straightforward approach of simply raising the bath temperature to its final value.

There are two conditions for this approach to be successful: first, the initial and final correlation lengths must
be sufficiently different; and second, the dynamics at the intermediate low temperature should be very fast
in order to make negligible the time necessary to match
$E_{\tw}[\mathcal{M}^2_{\mathrm{st}}] \approx \mathbb{E}^{T_{\text{f}}}[\mathcal{M}^2_{\mathrm{st}}]$.
The points in the process $\text{C}\to\text{A}\to\text{E}$ have been chosen accordingly.
In particular, the fast dynamics at $\text{A}$ is due to the its small correlation length (see~table~\ref{tab:xist}).

In the following discussion we use the notation
\begin{equation}\label{eq:chi-dyn-def}
\chidyn(t)=E_{\tw}[\mathcal{M}^2_{\mathrm{st}}]/N\,,
\end{equation}
although equation~\eqref{eq:chi-dyn-def} is a (useful) abuse of language because strictly speaking
the correspondence between the staggered susceptibility and $\mathcal{M}^2_{\mathrm{st}}$ holds
only at equilibrium (see equation~\eqref{eq:stagger-suscept-def}). 

\begin{figure}[t]
    \centering
    \includegraphics[width=0.48\linewidth]{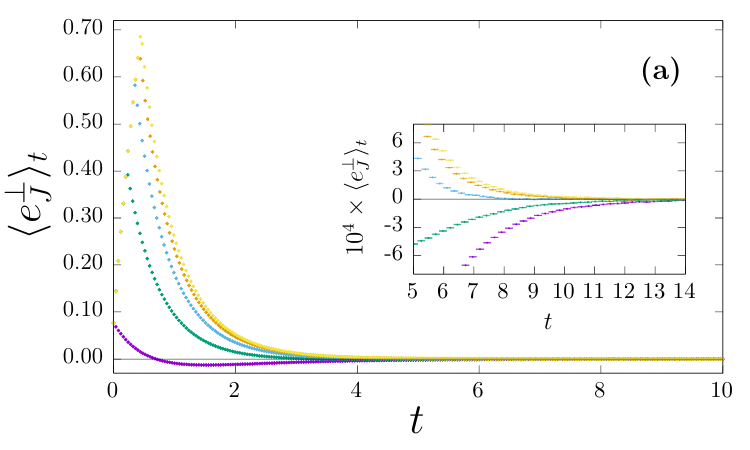}
    \hfill
    \includegraphics[width=0.48\linewidth]{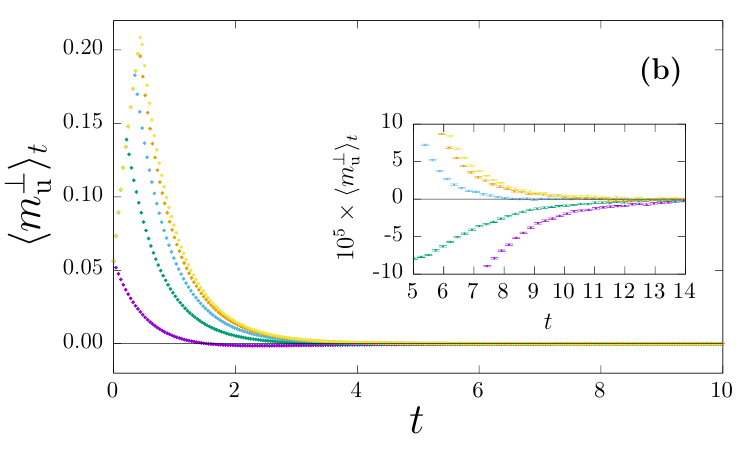}
    \vspace{0.5em}
    \includegraphics[width=0.48\linewidth]{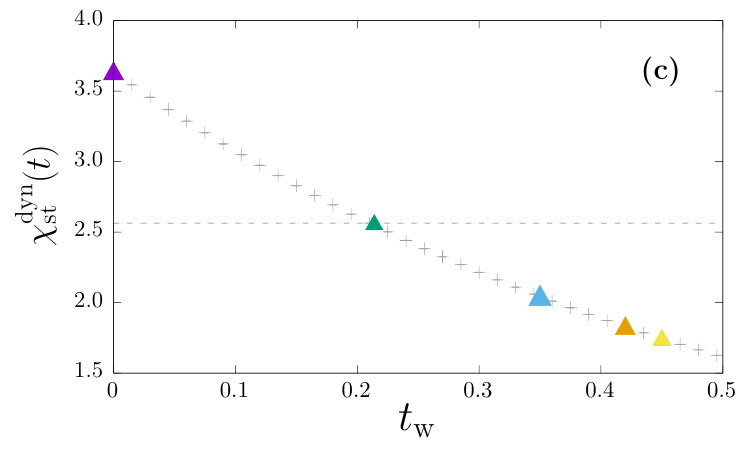}
    \hfill
    \includegraphics[width=0.48\linewidth]{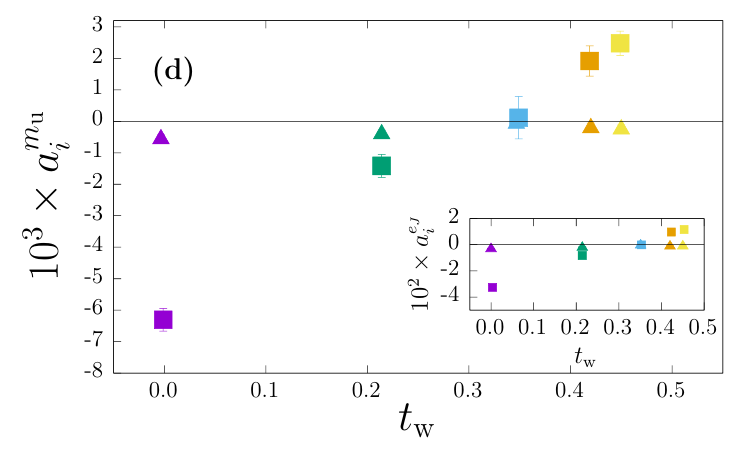}  
    \caption{Precooling protocol for different values of $t_{\mathrm{w}}$:
                 $t_{\mathrm{w}}$ = 0 (purple),
                 $t_{\mathrm{w}}$ = 0.214 (green),
                 $t_{\mathrm{w}}$ = 0.350 (blue),
                 $t_{\mathrm{w}}$ = 0.420 (orange),
                 $t_{\mathrm{w}}$ = 0.450 (yellow).
                 In panels (a) and (b) we show the relaxation to equilibrium of the exchange energy per spin and magnetization per spin, respectively.
                 Note that the fastest thermalization occurs for $t_{\mathrm{wait}}$ = 0.350. In (c) we show the evolution of $\chi^{\text{dyn}}_{\mathrm{st}}$
                 as a function of $t_{\mathrm{w}}$ with markers at each color-coded $t_{\mathrm{w}}$
                 (dashed red line: $\mathbb{E}^{T_{\mathrm{E}}}[\chi_{\mathrm{st}}]$ at the endpoint $\text{E}$ of the protocol).
                 In (d), we show the fitted amplitudes of the ansatz~\ref{eq:ansatz-1} with two exponential functions for the magnetization
                 (main panel) and exchange energy (inset: triangles: $a_{1}$, squares: $a_{2}$).
                 We have applied a small displacement in the horizontal axis to some of the points so that their error bars don't overlap.
                 Note that a clear zero-crossing occurs for $a_2$ as $t_{\mathrm{w}}$ is increased,
                 suggesting maximum mode-suppression at about $t_{\mathrm{w}}=0.350$.}
    \label{fig:precooling}
\end{figure}

These ideas are illustrated in figure~\ref{fig:precooling}. In particular, figure~\ref{fig:precooling}(c) compares
$\chidyn(\tw)$ with the target equilibrium value $\chi_{\text{st}}(\text{E})$ (dashed red line) for several values of $\tw$.
We see that the matching condition is met at $\tw=0.214$ and hence the exponential speed-up should be sought
for values near to $0.2$. The insets in figures~\ref{fig:precooling}(a) and (b) show that the approach to
equilibrium is from below at $\tw=0.214$ or from above at $\tw=0.420$. It is all too natural, then,
to expect that for an intermediate time $\tw$ an optimum  speed-up can be achieved. 

To be more precise we need to parametrize the approach to equilibrium, which,
as explained in section~\ref{sub:analysis_num_data}, is not trivial in the thermodynamic limit.
We choose the simplest approach in equation~\eqref{eq:ansatz-1} and fit $\mathcal{A}^\perp(t-\tw)$
to a sum of two exponential functions with the values for $\hat{\tau}_1$ and $\hat{\tau}_2$
quoted for point E in table~\ref{tab:compat_1}, for times $(t-\tw)> t_\text{min}=4.2$.\footnote{
The maximum time $t_{\text{max}}$ is fixed on a case-by-case basis using the three-standard deviate rule explained in section~\ref{sub:analysis_num_data}. The $p$-values of the fit are in the range  $[0.94281,\,1]$.}
The time shift $(t-\tw)$ is used because equation~\eqref{eq:ansatz-1} assumes that $t=0$ is the instant at
which the bath temperature reaches its final value. The amplitudes $a_j$ from the fits are shown in figure~\ref{fig:precooling}(d).
Whereas the amplitude $a_1$ is almost compatible with zero and essentially independent of $\tw$, the amplitude $a_2$
changes sign as $\tw$ increases. We may then look for a  special waiting time $\tw^*=0.350$ for which \emph{both}
amplitudes are compatible with zero. An important consistency check is that the two amplitudes vanish at $\tw^*$
for the two observables under consideration, namely $m_{\text{u}}$ and $e_J$.

Despite the effective nature of the description~\eqref{eq:ansatz-1}, the results of the preceding analysis are supported
by the strong speed-up displayed by the $\tw^*$-precooling protocol, see the cyan curves in figures~\ref{fig:precooling}(a) and (b).
Note that the figures display the total time duration of the precooling protocol,
including the overhead $\tw$ caused by the excursion to the lowest temperature. 

%%%%%%%%%%%%%%%%%%%%%%%%%%%%%%%%%%%%%%%%%%%%%%%%%%%%%%%%%%%%%%%%%%%%%%%%%%%%%%%%
%%%%%%%%%%%%%%%%%%%%%%%%%%%%%%%%%%%%%%%%%%%%%%%%%%%%%%%%%%%%%%%%%%%%%%%%%%%%%%%%

%%%%%%%%%%%%%%%%%%%%%%%%%%%%%%%%%%%%%%%%%%%%%%%%%%%%%%%%%%%%%%%%%%%%%%%%%%%%%%%%
\subsection{Direct and Inverse Mpemba effect}\label{sub:mpemba}
%%%%%%%%%%%%%%%%%%%%%%%%%%%%%%%%%%%%%%%%%%%%%%%%%%%%%%%%%%%%%%%%%%%%%%%%%%%%%%%%
As we explained in the Introduction, the direct Mpemba effect happens when the hotter of two otherwise identical systems
cools down faster than the colder system when both are placed in contact with a colder-than-both thermal bath.
The inverse Mpemba effect occurs when the colder of two systems heats up faster than the hotter system
when both are put in contact with a still hotter thermal bath.   

We discuss the Mpemba effect in two settings: by changing only the temperature and by changing both the temperature
and the magnetic field. The  coordinates in the phase diagram for the coldest thermal bath in our direct Mpemba effect
will be those of point B in  figure~\ref{fig:crit-suscept}(a), while the end point for the inverse Mpemba effect will be point E.

Again, we use equation~\eqref{eq:our-main-hyp} to select starting points that will show the Mpemba effect. 
What we need is that the left-had side of  equation~\eqref{eq:our-main-hyp}---as applied to the end-point---be 
non-zero for the protocol starting at the intermediate temperature, and be zero for the hottest of the 
two starting points. Since we only have access to the right-hand side of equation~\eqref{eq:our-main-hyp},
we need to choose an intermediate starting point with $\chi_{\text{st}}$ as different as possible to the final one. 
Of course, the other starting point should have $\chi_{\text{st}}$ identical to the final one. This argument will
give us a very good initial guess for the hottest starting point, that can be later refined to maximize the effect.

Given the above considerations, the natural candidates for the intermediate starting points 
are points D and C. Indeed, $\chi_{\text{st}}(\text{C})$ is as different from $\chi_{\text{st}}(\text{B})$ as 
possible without varying the magnetic field, while  $\chi_{\text{st}}(\text{D})$ differs even more at 
the price of varying $h$, see~\ref{fig:crit-suscept}(a) and (b). As for the hotter starting point, the preceding
criterion uniquely indicates point E.

\begin{figure}
    \centering
    \includegraphics[width=0.49\linewidth]{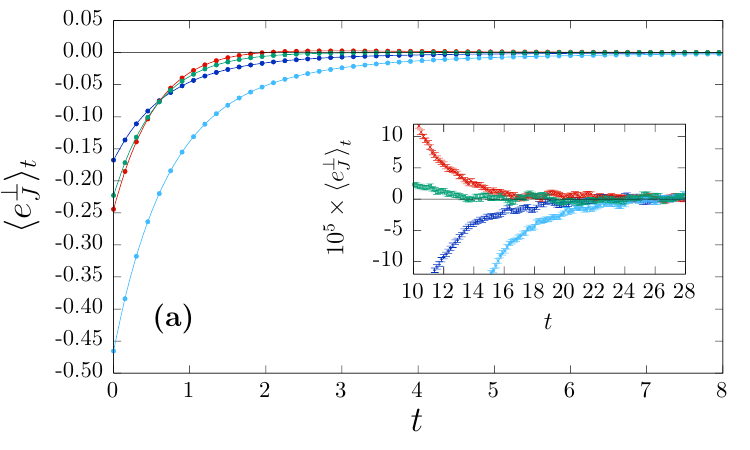}        
    \hfill
    \includegraphics[width=0.49\linewidth]{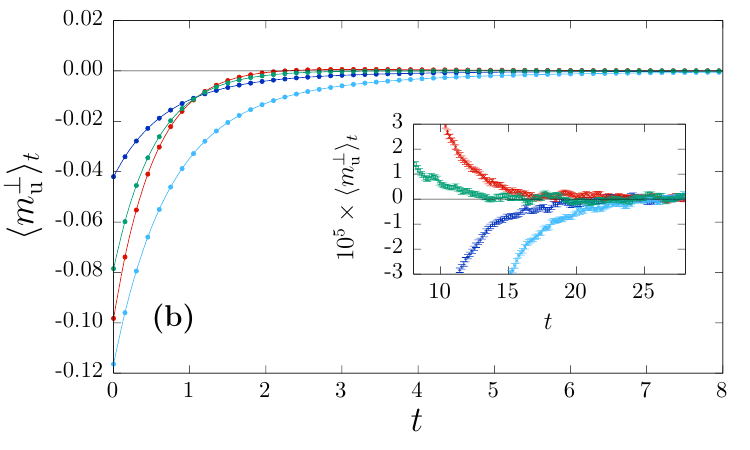}    
    \vspace{0.5em}
    \includegraphics[width=0.49\linewidth]{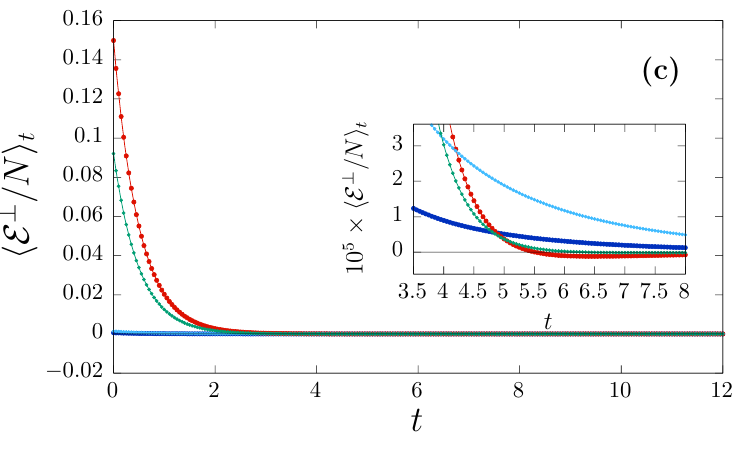}
    \hfill
    \includegraphics[width=0.49\linewidth]{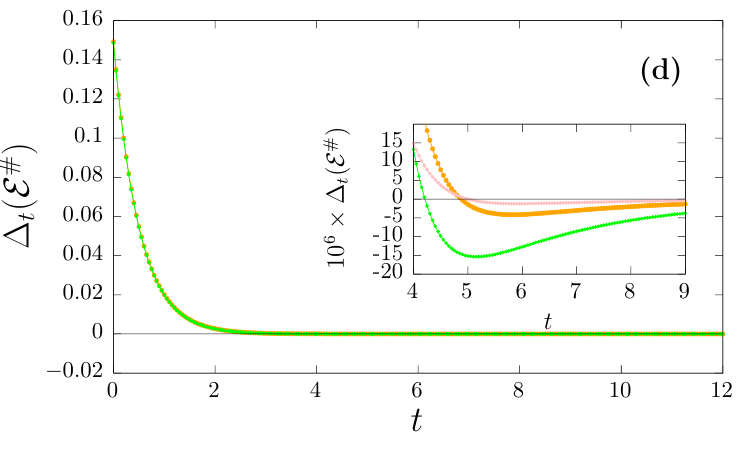}
    \vspace{0.5em}
    \includegraphics[width=0.49\linewidth]{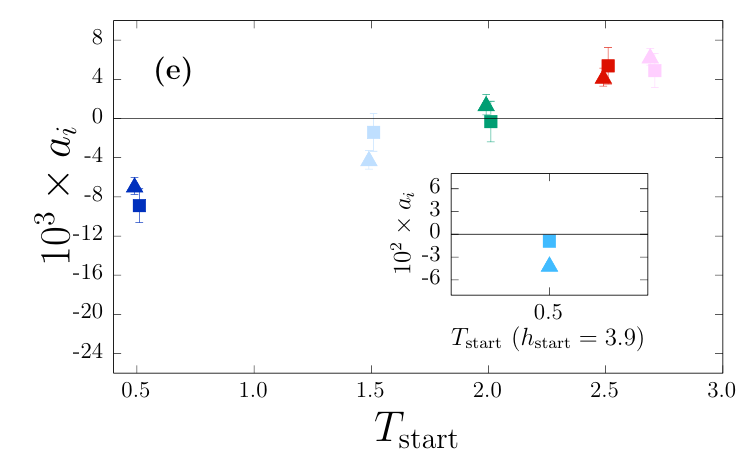}
    \hfill
    \includegraphics[width=0.49\linewidth]{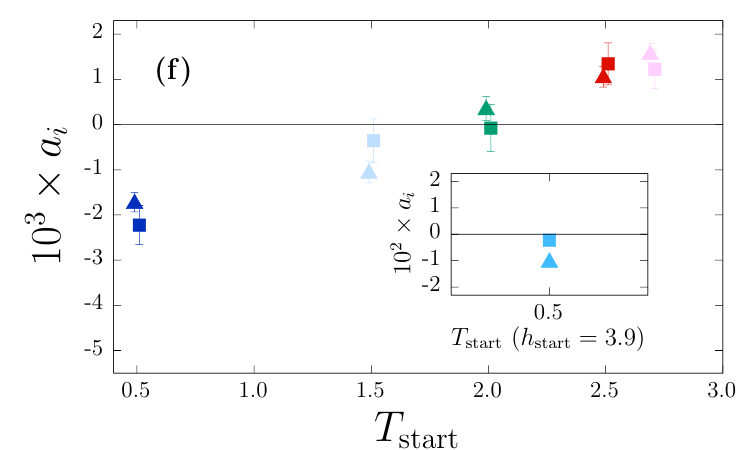}
    \caption{Direct Mpemba effects. In panels (a) to (d) we compare a fast path $\text{E} \to \text{B}$ (red)
                 vs two slower paths $\text{C} \to \text{B}$ (dark blue) and $\text{D} \to \text{B}$ (light blue)
                 and our fastest, near-optimal path $\text{F}=(T=2.0,h=4.01) \to \text{B}$ (olive green).
                 Note that point F is not in table~\ref{tab:working_points} because it has been found
                 specifically for the Mpemba protocol by iterative search using the amplitudes in panels (e) and (f).
                 The observables in the  different panels are:
                 (a) exchange energy per spin;
                 (b) magnetization per spin; and
                 (c) total energy per spin (lines are a cubic splines interpolation as a guide to eye).
                 Subtracted energies  $\Delta_t(\mathcal{E}^{\#})$ defined in equation~\eqref{eq:subtracted-direct} are shown in (d):
                  $\Delta_t(\mathcal{E}^{\text{C}\to\text{B}})$ (orange);
                  $\Delta_t(\mathcal{E}^{\text{D}\to\text{B}})$ (green) ; and
                  $\Delta_t(\mathcal{E}^{\text{F}\to\text{B}})$ (pink), with cubic spline interpolations as a guide to eye.
                  We show the amplitudes from fits to  ansatz~\eqref{eq:ansatz-1}---with two exponential functions---as a function of
                  $T_{\mathrm{start}}$  for the exchange energy (e) and  the magnetization (f) (triangles: amplitude $a_{1}$; squares: amplitudes $a_{2}$).
                  The additional translucent markers at $T_{\mathrm{start}}=1.5$ and $T_{\mathrm{start}}=2.7$, which do not correspond to curves
                  in previous panels, aid trend visualization. Data points have been slightly shifted horizontally ($a_1$ to the left, $a_2$ to the right)
                  to prevent the overlap of the error bars. A clear zero-crossing of both amplitudes occurs at about $T_{\mathrm{start}} = 2.0$.
                  Insets: Additional amplitudes for the process starting at point $\text{D}$, which includes both a jump in $h$ and in $T$.}
    \label{fig:mpemba_direct_crossh}
\end{figure}

Let us compare the above predictions with the results of our numerical simulations in figure~\ref{fig:mpemba_direct_crossh}.
We expect the $\text{E}\to\text{B}$ relaxation to be the fastest one (red points in the figure), $\text{C}\to\text{B}$
the intermediate (dark blue points), and $\text{D}\to\text{B}$ the slowest (cyan points).
Figure~\ref{fig:mpemba_direct_crossh} also shows  the results from a fourth protocol $\text{F}\to\text{B}$,
which is the fastest one because the starting point was obtained with the by now familiar optimization
protocol: we parametrize the relaxation through the effective model in equation~\eqref{eq:ansatz-1}
and select the starting temperature that causes the amplitudes to vanish, see figure~\ref{fig:mpemba_direct_crossh}(e) and (f).
These trends are very clear for the exchange energy (a) and the magnetization
(b), and much less clear for the energy density, where we have to resort to the inset in panel (c).  
We make the effect clearer by comparing
the time evolution of the total energies with the one for process $\text{E} \to \text{B}$
through the subtracted energies defined by
\begin{equation}\label{eq:subtracted-direct}
\Delta_t (\mathcal{E}^\#)=\frac{1}{N}\Big(E^{\text{E}\to\text{B}}_{t}[\mathcal{E}] \ -\ E^{\#}_{t}[\mathcal{E}]\Big)
\end{equation}
that are shown in figure~\ref{fig:mpemba_direct_crossh}(d) ($\#$ indicates the protocol that is being confronted with $\text{E}\to\text{B}$).
Although the energy for the fastest relaxation $\text{E}\to\text{B}$ is not the smallest at all times, it is certainly the most negative
during the final stages of the relaxation. Note that $\Delta_t(\mathcal{E}^{\text{F}\to\text{B}})$ becomes negative even though  $\text{F}\to\text{B}$
is our fastest process, because the final approach to equilibrium for the relaxation $\text{E}\to\text{B}$ is \emph{from below}, as shown in the inset to panel (c).

\begin{figure}
    \centering
    \includegraphics[width=0.49\textwidth]{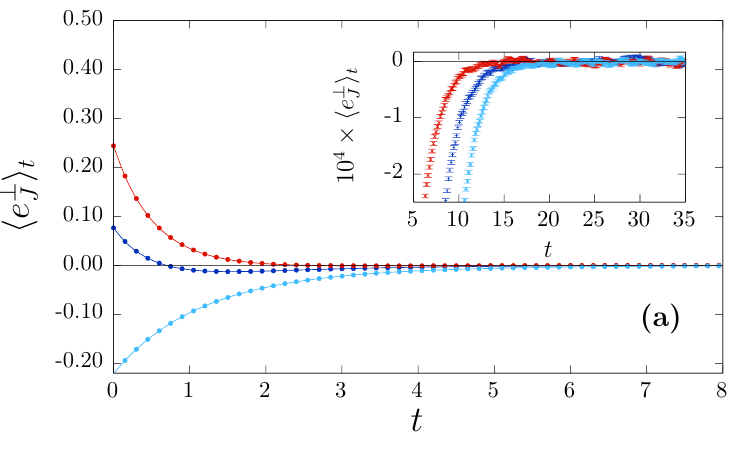}
    \hfill
    \includegraphics[width=0.49\textwidth]{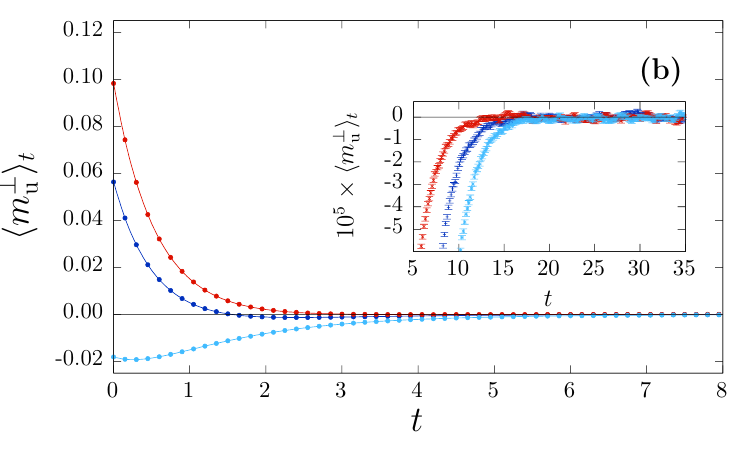}
    \vspace{0.5em}
    \includegraphics[width=0.49\textwidth]{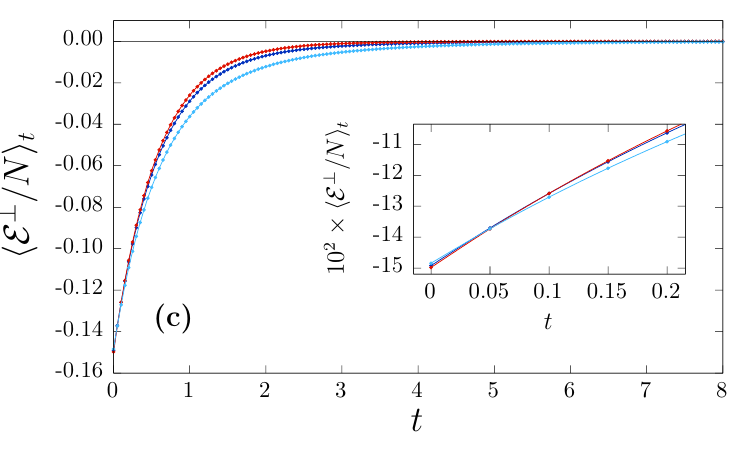}
    \hfill
    \includegraphics[width=0.49\textwidth]{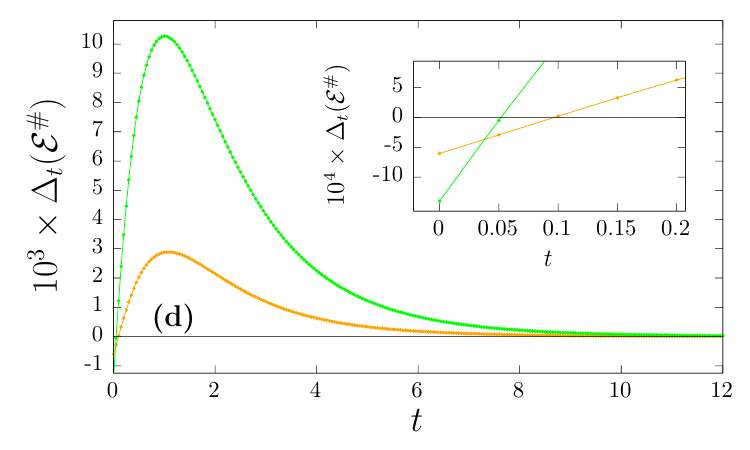}
    \vspace{0.5em}
    \includegraphics[width=0.49\linewidth]{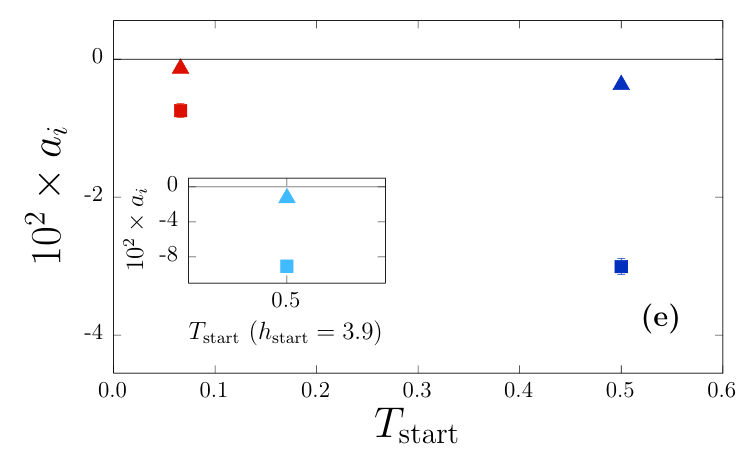}
     \hfill
    \includegraphics[width=0.49\linewidth]{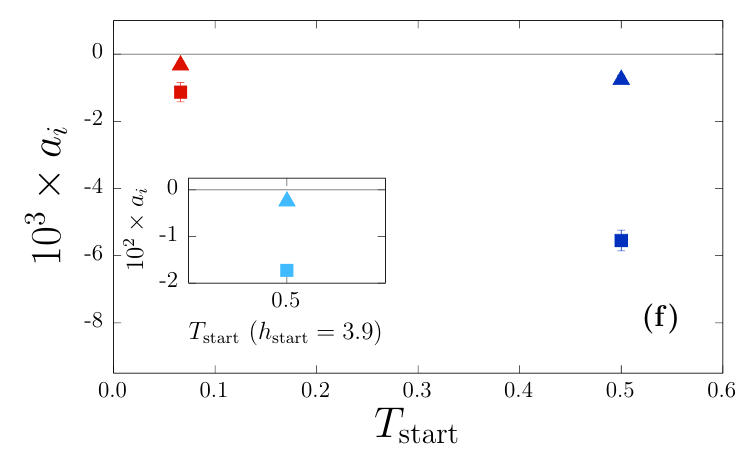}  
    \caption{Inverse Mpemba effects. In panels (a) to (d) we compare a faster path $\text{B} \to \text{E}$ (red)
                  vs two slower paths $\text{C} \to \text{E}$ (dark blue) and $\text{D} \to \text{E}$ (light blue). 
                  he magnitudes in the different panels are:
                  (a) exchange energy per spin;
                  (b)  magnetization per spin; and
                  (c) total energy per spin, with cubic splines interpolations as a guide to eye.
                  Subtracted energies  $\Delta_t(\mathcal{E}^{\#})$ defined in equation~\eqref{eq:subtracted-inverse} are shown in (d):
                  $\Delta_t(\mathcal{E}^{\text{C}\to\text{E}})$ (orange); and
                  $\Delta_t(\mathcal{E}^{\text{D}\to\text{E}})$ (green), with cubic spline interpolations as a guide to eye.
                  Notice that the light blue curves for $m_{\text{u}}$ and $e_{J}$ start the approach to equilibrium from below
                  while the other curves start from above, which is a consequence of the jump in $h$ that the light blue curve includes.
                  Nevertheless, the total energy behaves as expected.  We show the amplitudes from fits
                  to ansatz~\eqref{eq:ansatz-1}---with two exponential functions---as a function of $T_{\mathrm{start}}$
                  for (e) the exchange energy, and (f) the magnetization (triangles: amplitude $a_{1}$; squares: amplitudes $a_{2}$).
                  Insets: Additional amplitudes for the process starting at point $\text{D}$, which includes both a jump in $h$ and in $T$.}  
    \label{fig:mpemba_inverse_crossh}
\end{figure}

The inverse Mpemba effect is illustrated in Figure~\ref{fig:mpemba_inverse_crossh} using the same points B, C and D in the phase diagram.
Given their respective staggered susceptibilities in figure~\ref{fig:crit-suscept}, we expect $\text{B}\to\text{E}$ to be the fastest relaxation,
$\text{C}\to\text{E}$ to be slower, and $\text{D}\to\text{E}$ to be the slowest process. These expectations are confirmed for the exchange
energy (a) and the magnetization (b), although, at variance with the direct Mpemba effect, one needs to zoom in the final stages of the relaxation
to find the inverse effect because processes $\text{B}\to\text{E}$ and $\text{C}\to\text{E}$ undershoot the equilibrium value,
which is ultimately approached \emph{from below}. Again, cancellations arise that make the inverse Mpemba effect less clear,
although noticeable, for the energy density $\mathcal{E}$ (c). To  facilitate the comparison with process $\text{B}\to\text{E}$ we plot
in panel (d) the subtracted subtracted energies defined by
\begin{equation}\label{eq:subtracted-inverse}
\Delta_t (\mathcal{E}^\#)=\frac{1}{N}\Big( E^{\#}_{t}[\mathcal{E}]\ -\ E^{\text{B}\to\text{E}}_{t}[\mathcal{E}]\Big)\,.
\end{equation}
The fact that $\Delta_t(\mathcal{E})$ is positive in the final stages of the approach to equilibrium indicates that $\text{B}\to\text{E}$ is, indeed,
the fastest process. The reader may ask why we have not included an optimal curve analogous to the $\text{F}\to\text{B}$ relaxation
for the direct Mpemba effect. The reason can be inferred from the amplitudes for the fits to equation~\eqref{eq:ansatz-1} shown in panels (d) and (f):
linear extrapolations indicate that the optimal starting temperature, where the amplitudes vanish, would be \emph{negative}.
Therefore, from linear extrapolations of our data, it seems that we can achieve increasingly faster processes, but not an optimal one.
However, it is entirely possible that the evolution of the amplitudes with the starting temperature  would not be linear down to $T=0$,
in which case an optimal relaxation could be achieved.

%%%%%%%%%%%%%%%%%%%%%%%%%%%%%%%%%%%%%%%%%%%%%%%%%%%%%%%%%%%%%%%%%%%%%%%%%%%%%%%%
%%%%%%%%%%%%%%%%%%%%%%%%%%%%%%%%%%%%%%%%%%%%%%%%%%%%%%%%%%%%%%%%%%%%%%%%%%%%%%%%

%%%%%%%%%%%%%%%%%%%%%%%%%%%%%%%%%%%%%%%%%%%%%%%%%%%%%%%%%%%%%%%%%%%%%%%%%%%%%%%%
%%%%%%%%%%%%%%%%%%%%%%%%%%%%%%%%%%%%%%%%%%%%%%%%%%%%%%%%%%%%%%%%%%%%%%%%%%%%%%%%

%%%%%%%%%%%%%%%%%%%%%%%%%%%%%%%%%%%%%%%%%%%%%%%%%%%%%%%%%%%%%%%%%%%%%%%%%%%%%%%%
\section{Conclusions}\label{sec:conclusions}
%%%%%%%%%%%%%%%%%%%%%%%%%%%%%%%%%%%%%%%%%%%%%%%%%%%%%%%%%%%%%%%%%%%%%%%%%%%%%%%%
This work started with a very disappointing realization: the naive expectation about how the thermodynamic limit for
relaxation processes occurs is untenable. An extensive quantity  $\mathcal{A}$ with a variance displaying the standard scaling,
\begin{equation}
    \mathbb{E}^{T_{\text{f}}}[\mathcal{A}]\propto N\,,\quad \mathbb{E}^{T_{\text{f}}}[\mathcal{A}^2]-\mathbb{E}^{T_{\text{f}}}[\mathcal{A}]^2\propto N\,,
\end{equation}
has a relaxation that for finite $N$ is a sum of $2^N-1$ exponentially decaying terms. Unfortunately, the leading term in this expansion does 
not have a one-to-one correspondence to an allegedly leading term in the thermodynamic limit. The natural conjecture is that in the limit $N\to\infty$
the discrete sum goes over to a continuous distribution (see equation~\eqref{eq:evolution-intensive}). Therefore, it is unclear how the 
standard mathematical discussion of Mpemba-like anomalous effects for small $N$ systems~\cite{TB26} carries over to the thermodynamic limit,
since this standard discussion relies on devising protocols that cause the amplitude for this (non-existing for large systems) leading term 
to vanish.

As a most natural way out of this impasse, we have resorted to the ansatz expressed in equation~\eqref{eq:our-main-hyp}: the spectral projector onto 
the slowest relaxation times in the distribution is related to the fluctuations of the order parameter for the metastable phase. This idea 
goes back to Parisi~\cite{PA88} and, ultimately, to Langer~\cite{LA67}. This ansatz is expected to  be accurate in regions of the phase
diagram where two phases coexist, and is the natural extension to the thermodynamic limit of the ansatz put forth for finite systems
in reference~\cite{GL24}. Our conjecture, when combined with the non-monotonic temperature evolution of the susceptibility associated
to the metastable phase (which is a generic feature), leads naturally to a qualitative (and even semi-quantitative)
prediction of several Mpemba-like anomalous relaxations.

These ideas are illustrated in the context of the 2D antiferromagnetic Ising model with an externally applied magnetic field. The phase diagram 
contains a staggered phase and a paramagnetic phase. We choose to work in the paramagnetic phase (therefore, the metastable phase
is the staggered  one) but near the phase-transition line. Hence, the staggered susceptibility $\chi_{\text{st}}$ is the relevant magnitude
for our purposes. The  presence of a critical point at zero field implies that we can have arbitrarily large values of $\chi_{\text{st}}$,
with arbitrarily large  values of the corresponding correlation length. This added flexibility as compared with previous known examples
leads to consider protocols  where not only the temperature, but also the magnetic field are varied. In this way we achieve variations
of  $\chi_{\text{st}}$ much larger than the maximum value attainable in previous works.

Given the above framework, a simple glance at the susceptibility permits us to predict protocols where the direct and inverse Mpemba effects 
are realized, including extensions of the standard protocols where not only the temperature, but also the magnetic field, are varied. Related 
effects, such as faster heating through pre-cooling and asymmetries between heating and cooling are also easily predicted. We check 
that the thermodynamic limit is realized in our simulations (in the sense that residual $N$-dependencies are smaller than our rather small 
statistical error). Although our conjecture does not lead to the optimal protocol where the maximum speed-up is achieved, it provides a fairly 
reasonable first guess, allowing us to refine the protocol with a straightforward search.

Since phase-coexisting regions are fairly common in condensed matter physics, we could expect that the method developed in this paper could be 
applied to other systems. However, the main difficulty will be finding a quantity playing the same role of our staggered magnetization in these systems. For 
instance, a simple liquid-vapor equilibrium would not do because there is not a different order parameter for each of the two coexisting 
phases.  Indeed, the non-monotonic temperature evolution of our staggered susceptibility stems from the fact that the staggered magnetization 
vanishes at both $T=0$ and $T\to\infty$. Thus, we can add having separated order parameters for each of the coexisting phases as a second requirement.
%%%%%%%%%%%%%%%%%%%%%%%%%%%%%%%%%%%%%%%%%%%%%%%%%%%%%%%%%%%%%%%%%%%%%%%%%%%%%%%%
\funding{The research was supported by Ministerio de Ciencia, Innovaci\'on y Universidades (MICIU, Spain),
              Agencia Estatal de Investigaci\'on (AEI, Spain, MCIN/AEI/10.13039/501100011033),
              and European Regional Development Fund (ERDF, A way of making Europe) through
              Grant nos.~PID2022-136374NB-C21 and PID2024-155527NB-I00;
              by Consejer\'{\i}a de Universidad, Investigaci\'on e Innovaci\'on and by ERDF Andalusia Program 2021-2027
              through Grant no.~C-EXP-251-UGR23;
              and by Universidad Complutense de Madrid through Grant no.~ PR12/24-31565.}
%%%%%%%%%%%%%%%%%%%%%%%%%%%%%%%%%%%%%%%%%%%%%%%%%%%%%%%%%%%%%%%%%%%%%%%%%%%%%%%%
\bibliographystyle{unsrtnat}
\bibliography{anomrelax}
%\bibliography{mpemba}
%%%%%%%%%%%%%%%%%%%%%%%%%%%%%%%%%%%%%%%%%%%%%%%%%%%%%%%%%%%%%%%%%%%%%%%%%%%%%%%%
\end{document}